\documentclass[11pt]{article}

\usepackage{latexsym}
\usepackage{psfig}

\newcommand{\an}{\mbox{{\it 2N}}}
\newcommand{\bn}{\mbox{{\it 3N}}}
\newcommand{\tn}{\mbox{{\it tN}}}
\newcommand{\tnm}{\mbox{{\it tNM}}}
\newcommand{\anm}{\mbox{{\it 2NM}}}
\newcommand{\bnm}{\mbox{{\it 3NM}}}
\newcommand{\tna}{\mbox{{\it tNA}}}
\newcommand{\ana}{\mbox{{\it 2NA}}}
\newcommand{\bna}{\mbox{{\it 3NA}}}
\newcommand{\At}{\mbox{{\it At}}}
\newcommand{\WS}{\mbox{{\it WS}}}
\newcommand{\SPM}{{\cal SP}}
\newcommand{\STM}{{\cal ST}}

\newcommand{\n}{\mbox{{\bf not}}}

\newcommand{\lm}{\mbox{{\it lm}}}

\newcommand{\comp}{\mbox{{\it comp}}}

\newcommand{\lar}{\leftarrow}

\newtheorem{theorem}{Theorem}[section]
\newtheorem{lemma}[theorem]{Lemma}

\newtheorem{corollary}[theorem]{Corollary}

\newtheorem{definition}[theorem]{Definition}

\begin{document}
\ \\
\begin{center}
{\large\bf Fixed-parameter complexity of semantics for logic
programs\footnote{A preliminary version of this paper
         appeared in the Proceedings of ICLP 2001 published by 
         Springer Verlag. }}\\
\ \\
\ \\
Zbigniew Lonc\footnote{On leave from Faculty of Mathematics and
Information Science, Warsaw University of Technology.} and 
Miros\l aw Truszczy\'nski\\
\ \\
Department of Computer Science\\
University of Kentucky\\
Lexington KY 40506-0046, USA\\
\tt{lonc|mirek@cs.engr.uky.edu}
\end{center}

\begin{abstract}
A decision problem is called {\em parameterized} if its input is a pair
of strings. One of these strings is referred to as a {\em parameter}.
The problem: given a propositional logic program $P$ and a non-negative
integer $k$, decide whether $P$ has a stable model of size no more than
$k$, is an example of a parameterized decision problem with $k$ serving
as a parameter. Parameterized problems that are NP-complete often become
solvable in polynomial time if the parameter is fixed. The problem to
decide whether a program $P$ has a stable model of size no more than
$k$, where $k$ is fixed and not a part of input, can be solved in time 
$O(mn^k)$, where $m$ is the size of $P$ and $n$ is the number of atoms
in $P$. Thus, this problem is in the class P. However, algorithms with
the running time given by a polynomial of order $k$ are not satisfactory
even for relatively small values of $k$. 

The key question then is whether significantly better algorithms (with 
the degree of the polynomial not dependent on $k$) exist. To tackle it,
we use the framework of fixed-parameter complexity. We establish the
fixed-parameter complexity for several parameterized decision problems 
involving models, supported models and stable models of logic programs. 
We also establish the fixed-parameter complexity for variants of these 
problems resulting from restricting attention to Horn programs and to 
purely negative programs. Most of the problems considered in the paper
have high fixed-parameter complexity. Thus, it is unlikely that fixing 
bounds on models (supported models, stable models) will lead to fast
algorithms to decide the existence of such models.
\end{abstract}

\section{Introduction}
\label{intro}

In this paper we study the complexity of parameterized decision problems 
concerning models, supported models and stable models of logic programs. 
In our investigations, we use the framework of the {\em fixed-parameter 
complexity} introduced by Downey and Fellows \cite{df97}. This framework 
was previously used to study the problem of the existence of stable models 
of logic programs in \cite{tru99j}. Our present work extends results obtained 
there. First, in addition to the class of all finite propositional logic 
programs, we consider its two important subclasses: the class of Horn 
programs and the class of purely negative programs. Second, in addition 
to stable models of logic programs, we also study supported models and 
arbitrary models. 

A decision problem is called {\em parameterized} if its inputs are 
{\em pairs} of items. The second item in a pair is referred to as 
a {\em parameter}. The problem to decide, given a logic program $P$ and 
an integer $k$, whether $P$ has a stable model with {\em at most} $k$ 
atoms is an example of a parameterized decision problem. This
parameterized problem is NP-complete. However, fixing $k$ (in other words, 
$k$ is no longer regarded as a part of the input) makes the problems 
simpler. It becomes solvable in polynomial time. The following 
straightforward algorithm works: for every subset $M \subseteq \At(P)$ 
of cardinality at most $k$, check whether $M$ is a stable model of $P$. 
The check can be implemented to run in linear time in the size of the 
program. If $n$ stands for the number of atoms in $P$, there are $O(n^k)$ 
sets to be tested. Thus, the overall running time of this algorithm is 
$O(mn^k)$, where $m$ is the size of the input program $P$. This 
discussion also applies to analogous problems in logic programming 
concerned with the existence of models and supported models.


Unfortunately, algorithms with running times given by $O(mn^k)$
are not practical even for quite small values of $k$. The question then
arises whether better algorithms can be found, for instance, algorithms 
whose running-time estimate would be given by a polynomial of the order 
that {\em does not depend on} $k$. Such algorithms, if they existed, 
could be practical for a wide range of values of $k$ and could find 
applications in computing stable models of logic programs. 

This question is the subject of our work. We also consider similar 
questions concerning related problems of deciding the existence of models, 
supported models and stable models of cardinality {\em exactly} $k$ and
{\em at least} $k$. We refer to all these problems as {\em small-bound} 
problems since $k$, when fixed, can be regarded as ``small''
($\frac{k}{|{\footnotesize\At}(P)|}$ converges to 0 as $|\At(P)|$ goes to infinity).
In addition, we study problems of existence of models, supported models and 
stable models of cardinality at most $|\At(P)|-k$, exactly $|\At(P)|-k$ 
and at least $|\At(P)|-k$. We refer to these problems as {\em 
large-bound} problems, since $|\At(P)|-k$, for a fixed $k$, can be thought 
of as ``large'' ($\frac{|{\footnotesize\At}(P)|-k}{|{\footnotesize\At}(P)|}$ converges to 1 
as $|\At(P)|$ goes to infinity).

We address these questions using the framework of fixed-parameter 
complexity \cite{df97}. Most of our results are 
negative. They provide strong evidence that for many parameterized problems 
considered in the paper there are no algorithms whose running time could 
be estimated by a polynomial of order independent of $k$. 

Formally, a {\em parameterized} decision problem is a set $L\subseteq 
\Sigma^*\times\Sigma^*$, where $\Sigma$ is a fixed alphabet. By selecting 
a concrete value $\alpha\in \Sigma^*$ of the parameter, a parameterized 
decision problem $L$ gives rise to an associated {\em fixed-parameter} 
problem $L_\alpha =\{x:(x,\alpha)\in L\}$.

A parameterized problem $L\subseteq \Sigma^*\times\Sigma^*$ is 
{\em fixed-parameter tractable} if there exist a constant $t$, an 
integer function $f$ and an algorithm $A$ such that $A$ determines 
whether $(x,y)\in L$ in time $f(|y|)|x|^t$ ($|z|$ stands for the 
length of a string $z\in \Sigma^*$). We denote the class of 
fixed-parameter tractable problems by FPT. Clearly, if a parameterized 
problem $L$ is in FPT, then each of the associated fixed-parameter 
problems $L_y$ is solvable in polynomial time by an algorithm whose 
exponent does not depend on the value of the parameter $y$. Parameterized
problems that are not fixed-parameter tractable are called {\em
fixed-parameter intractable}.

To study and compare the complexity of parameterized problems Downey and 
Fellows proposed the following notion of {\em fixed-parameter
reducibility} (or, simply, {\em reducibility}). 

\begin{definition} \label{red}
A parameterized problem $L$ can be {\em reduced} to a parameterized 
problem $L'$ if there exist a constant $p$, an integer function $q$, and 
an algorithm $A$ such that:
\begin{enumerate}
\item \label{red-1}
$A$ assigns to each instance $(x,y)$ of $L$ an instance $(x',y')$ 
of $L'$,
\item \label{red-2}$A$ runs in time $O(q(|y|)|x|^p)$,
\item \label{red-3}$x'$ depends upon $x$ and $y$, and $y'$ depends 
upon $y$ only,
\item \label{red-4} $(x,y)\in L$ if and only if $(x',y')\in L'$.
\end{enumerate}
\end{definition}
We will use this notion of reducibility throughout the paper. If for
two parameterized problems $L_1$ and $L_2$, $L_1$ can be reduced to $L_2$
and conversely, we say that $L_1$ and $L_2$ are {\em fixed-parameter
equivalent} or, simply, {\em equivalent}.

Downey and Fellows \cite{df97} defined a hierarchy of complexity classes 
called the {\em W hierarchy}:
\begin{equation}\label{eq1}
{\rm FPT} \subseteq {\rm W[1]} \subseteq {\rm W[2]} \subseteq {\rm W[3]}
\subseteq \ldots \ .
\end{equation}
The classes W[$t$] can be described in terms of problems that are complete
for them (a problem $D$ is {\em complete} for a complexity class $\cal
E$ if $D\in{\cal E}$ and every problem in this class can be reduced to 
$D$). Let us call a Boolean formula {\em $t$-normalized} if it is of the 
form of conjunction-of-disjunctions-of-conjunctions ... of literals, 
with $t$ being the number of conjunctions-of, disjunctions-of expressions 
in this definition. For example, 2-normalized formulas are conjunctions
of disjunctions of literals. Thus, the class of 2-normalized formulas is 
precisely 
the class of CNF formulas. We define the {\em weighted $t$-normalized
satisfiability problem} as:
\begin{description}
\item[$\WS(t)$:] Given a $t$-normalized formula $\Phi$ and a
non-negative integer $k$, decide whether there is a model of $\Phi$ 
with exactly $k$ atoms (or, alternatively, decide whether there is a 
satisfying valuation for $\Phi$ which assigns the logical value {\bf 
true} to exactly $k$ atoms).
\end{description}
Downey and Fellows show that for every $t\geq 2$, the problem 
$\WS(t)$ is complete for the class W[$t$]. They also show that 
a restricted version of the problem $\WS(2)$:
\begin{description}
\item[$\WS_2(2)$:] Given a 2-normalized formula $\Phi$ with each clause
consisting of at most two literals, and an integer $k$,
decide whether there is a model of $\Phi$ with exactly $k$
atoms
\end{description} 
is complete for the class W[1]. There is strong evidence
suggesting
that all the implications in (\ref{eq1}) are proper. Thus, proving 
that a parameterized problem is complete for a class W[t], $t\geq 1$, is 
a strong indication that the problem is not fixed-parameter tractable. 

As we stated earlier, in the paper we study the complexity of 
parameterized problems related to logic programming. All these problems 
ask whether an input program $P$ has a model, supported model or a
stable model satisfying some cardinality constraints involving another 
input parameter, an integer $k$. They can be categorized into two general 
families: {\em small-bound} problems and {\em large-bound} problems. In 
the formal definitions given below, $\cal C$ denotes a class of logic 
programs, ${\cal D}$ represents a class of models of interest and 
$\Delta$ stands for one of the three arithmetic relations: ``$\leq$'', 
``$=$'' and ``$\geq$''.  
 
\begin{description} 
\item[${\cal D}_\Delta({\cal C})$:] Given a logic program $P$ from class
$\cal C$ and an integer $k$, decide whether $P$ has a model $M$ from
class $\cal D$ such that $|M|\ \Delta\ k$.
\item[${\cal D}'_\Delta({\cal C})$:] Given a logic program $P$ from class
$\cal C$ and an integer $k$, decide whether $P$ has a model $M$ from
class $\cal D$ such that $(|\At(P)|-k)\ \Delta\ |M|$. 
\end{description}

In the paper, we consider three classes of programs: the class of Horn 
programs ${\cal H}$, the class of purely negative programs ${\cal N}$, 
and the class of all programs ${\cal A}$. We also consider three classes 
of models: the class of all models ${\cal M}$, the class of supported 
models ${\cal SP}$ and the class of stable models ${\cal ST}$. 

Thus, for example, the problem ${\cal SP}_\leq({\cal N})$ asks whether a 
purely negative logic program $P$ has a supported model $M$ with no 
more than $k$ atoms ($|M|\leq k$). The problem ${\cal ST}'_\leq({\cal A})$ 
asks whether a logic program $P$ (with no syntactic restrictions) has a 
stable model $M$ in 
which at most $k$ atoms are false ($|\At(P)|-k \leq |M|$). Similarly, the 
problem ${\cal M}'_\geq({\cal H})$ asks whether a Horn program $P$ has 
a model $M$ in which at least $k$ atoms are false ($|\At(P)|-k \geq |M|$). 

In the three examples given above and, in general, for all problems
${\cal D}_\Delta({\cal C})$ and ${\cal D}'_\Delta({\cal C})$, the input
instance consists of a logic program $P$ from the class ${\cal C}$ and of
an integer $k$. We will regard these problems as parameterized with $k$. 
Fixing $k$ (that is, $k$ is no longer a part of input but an element of 
the problem description) leads to the fixed-parameter versions of these 
problems. We will denote them ${\cal D}_\Delta({\cal C},k)$ and 
${\cal D}'_\Delta({\cal C},k)$, respectively. 

In the paper, for all but three problems ${\cal D}_\Delta({\cal C})$ and
${\cal D}'_\Delta({\cal C})$, we establish their fixed-parameter 
complexities. Our results are summarized in Tables \ref{tab1} - \ref{tab3}.

\begin{table}[h]
\begin{center}
\begin{tabular}{|c||c|c|c|}
\hline
            & $\cal H$  & ${\cal N}$& ${\cal A}$ \\
\hline\hline
${\cal M}$  & P         & P         & P \\
\hline
${\cal M}'$ & P         & W[1]-c    & NP-c \\
\hline
$\SPM$      & P         & NP-c      & NP-c \\
\hline
$\SPM'$     & P         & NP-c      & NP-c \\
\hline
$\STM$      & P         & NP-c      & NP-c \\
\hline
$\STM'$     & P         & NP-c      & NP-c \\
\hline
\end{tabular}
\caption{The complexities of the problems ${\cal D}_\geq({\cal C})$ and
${\cal D}'_\geq({\cal C})$.}
\label{tab1}
\end{center}
\end{table}

In Table \ref{tab1}, we list the complexities of all problems in which
$\Delta = \mbox{``$\geq$''}$. 
Small-bound problems of this type ask about the existence of models of 
a program $P$ that contain at least $k$ atoms. Large-bound problems in 
this group are concerned with the existence of models that contain at 
most $|\At(P)|-k$ atoms (the number of false atoms in these models is
at least $k$). 
From the point of view of the fixed-parameter complexity, these
problems are 
not very interesting. Several of them remain NP-complete even when $k$ 
is fixed. In other words, fixing $k$ does not simplify them enough to 
make them tractable. For this reason, all the entries in Table \ref{tab1}, 
listing the complexity as NP-complete (denoted by NP-c in the table), 
refer to fixed-parameter versions ${\cal D}_\geq({\cal C},k)$ and 
${\cal D}'_\geq({\cal C},k)$ of problems ${\cal D}_\geq({\cal C})$ and 
${\cal D}'_\geq({\cal C})$. The problem ${\cal M}'_\geq({\cal A},k)$
is NP-complete for every fixed $k\geq 1$. All other fixed-parameter
problems in Table \ref{tab1} that are marked NP-complete are NP-complete 
for every value $k\geq 0$. 

On the other hand, many problems ${\cal D}_\geq({\cal C})$ and
${\cal D}'_\geq({\cal C})$ are ``easy''. They are fixed-parameter
tractable in a strong sense. They can be solved in polynomial time even 
{\em without} fixing $k$. This is indicated by marking the corresponding
entries in Table \ref{tab1} with P (for the class P) rather than with 
FPT. There is only one exception, the problem ${\cal M}'_\geq({\cal N})$, 
which is W[1]-complete. 

Small-bound problems for the cases when $\Delta = \mbox{``$=$'' or 
``$\leq$''}$ can be viewed as problems of deciding the existence of 
``small'' models, that is, models containing exactly $k$ or at most $k$
atoms. Indeed, for a fixed $k$ and the number of atoms in a program
going to infinity, the ratio of the number of true atoms to the number 
of all atoms converges to 0 ($k$ is ``small'' with respect to
$|\At(P)|$). The fixed-parameter complexities of these 
problems are summarized in Table \ref{tab2}.

\begin{table}[h]
\begin{center}
\begin{tabular}{|c||c|c|c|c|c|c|}
\hline
 & ${\cal H}_\leq$ & ${\cal H}_=$ & ${\cal N}_\leq$ & ${\cal N}_=$ &
${\cal A}_\leq$ & ${\cal A}_=$ \\
\hline\hline
${\cal M}$ & P & W[1]-c & W[2]-c & W[2]-c & W[2]-c & W[2]-c \\
\hline
$\SPM$ & P & W[1]-h, & W[2]-c & W[2]-c & W[2]-c & W[2]-c \\
 & & in W[2] & & & & \\
\hline
$\STM$ & P & P & W[2]-c & W[2]-c & W[2]-c & W[2]-c \\
\hline
\end{tabular}
\caption{The complexities of the problem of computing small models
(small-bound problems, the cases of $\Delta= \mbox{``$=$''}$ and 
``$\leq$'').}
\label{tab2}
\end{center}
\end{table}

The problems involving the class of all purely negative programs and the 
class of all programs are W[2]-complete. This is a strong indication
that they are fixed-parameter intractable. All problems of the form 
${\cal D}_\leq({\cal H})$ are 
fixed-parameter tractable. In fact, they are solvable in polynomial time 
even without fixing the parameter $k$. We indicate this by marking the 
corresponding entries with P. Similarly, the problem $\STM_=({\cal H})$ 
of deciding whether a Horn logic program $P$ has a stable model of size 
exactly $k$ is in P. However, perhaps somewhat surprisingly, the
remaining two problems involving Horn logic programs and
$\Delta=\mbox{``$=$''}$ are harder. We proved that the problem 
${\cal M}_=({\cal H})$ is W[1]-complete and that the problem 
$\SPM_=({\cal H})$ is W[1]-hard. Thus, they most likely are not
fixed-parameter tractable. We also showed that the problem
$\SPM_=({\cal H})$ is in the class W[2]. The exact 
fixed-parameter complexity of $\SPM_=({\cal H})$ remains unresolved. 

Large-bound problems for the cases when $\Delta = \mbox{``$=$'' or
``$\leq$''}$ can be viewed as problems of deciding the existence of
``large'' models, that is, models with a small number of false atoms
--- equal to $k$ or less than or equal to $k$.  Indeed, for a fixed $k$
and the number of atoms in a program going to infinity, the ratio of the 
number of true atoms to the number of all atoms converges to 1 ($k$ is
``large'' with respect to $|\At(P)|$).
The fixed-parameter complexities of these problems are summarized in 
Table \ref{tab3}.

\begin{table}[h]
\begin{center}
\begin{tabular}{|c||c|c|c|c|c|c|}
\hline
 & ${\cal H}_\leq$ & ${\cal H}_=$ & ${\cal N}_\leq$ & ${\cal N}_=$ &
${\cal A}_\leq$ & ${\cal A}_=$ \\
\hline\hline
${\cal M}'$ & P & W[2]-c & P & W[1]-c & P & W[2]-c \\
\hline
$\SPM'$ & P & W[3]-c, & W[2]-c & W[2]-c & W[3]-c & W[3]-c \\
\hline
$\STM'$& P & P & W[2]-c & W[2]-c & W[3]-h & W[3]-h \\
\hline
\end{tabular}
\caption{The complexities of the problems of computing large models 
(large-bound problems, the cases of $\Delta= \mbox{``$=$''}$ and
``$\leq$'').}
\label{tab3}
\end{center}
\end{table}

The problems specified by $\Delta = \mbox{``$\leq$''}$ and concerning 
the existence of models are in P. Similarly, the problems specified by
$\Delta = \mbox{``$\leq$''}$ and involving Horn programs are solvable
in polynomial time. Lastly, the problem $\STM'_=({\cal H})$ is in P, as
well. These problems are in P even without fixing $k$ and 
eliminating it from input. All other problems in this group have higher
complexity and, in all likelihood, are fixed-parameter intractable. One 
of the problems, ${\cal M}'_=({\cal N})$, is W[1]-complete. Most of the 
remaining problems are W[2]-complete. Surprisingly, some problems are 
even harder. Three problems concerning supported models are W[3]-complete. 
For two problems involving stable models, $\STM'_=({\cal A})$ and 
$\STM'_\leq({\cal A})$, we could only prove that they are W[3]-hard. For 
these two problems we did not succeed in establishing any upper bound 
on their fixed-parameter complexities. 

The study of fixed-parameter tractability of problems occurring in the
area of nonmonotonic reasoning is a relatively new research topic.
The only two other papers we are aware of are \cite{tru99j} and
\cite{gss99}. The first of these two papers provided a direct motivation
for our work here (we discussed it earlier). In the second one, the 
authors focused on parameters describing {\em structural} properties 
of programs. They showed that under some choices of the parameters decision 
problems for nonmonotonic reasoning become fixed-parameter tractable.

Our results concerning computing stable and supported models for logic
programs are mostly negative. Parameterizing basic decision problems by
constraining the size of models of interest does not lead (in most
cases) to fixed-parameter tractability. 

There are, however, several interesting aspects to our work. First, we 
identified some problems that are W[3]-complete or W[3]-hard. 
Relatively few problems from these classes were known up to now \cite{df97}. 
Second, in the context of the polynomial hierarchy, there is no distinction
between the problem of existence of models of specified sizes of
clausal propositional theories and similar problems concerning models,
supported models and stable models of logic programs. All these
problems are NP-complete. However, when we look at the complexity of
these problems in a more detailed way, from the perspective of
fixed-parameter complexity, the equivalence is lost. Some problems are 
W[3]-hard, while problems concerning existence of models of
2-normalized formulas are W[2]-complete or easier. Third, our results 
show that in the context of fixed-parameter tractability, several 
problems involving models and supported models are hard even for the 
class of Horn programs. Finally, our work leaves three problems 
unresolved. While we obtained some bounds for the problems 
$\SPM_=({\cal H})$, $\STM'_\leq({\cal A})$ and $\STM'_=({\cal A})$, we 
did not succeed in establishing their precise fixed-parameter complexities.

The rest of our paper is organized as follows. In the next section, we
review relevant concepts in logic programming. Next, we present several 
useful fixed-parameter complexity results for problems of the existence of 
models for propositional theories of certain special types. We also
state and prove there some auxiliary results on the hardness of problems 
concerning the existence of stable and supported models. We study the 
complexity of the problems ${\cal D}_\geq({\cal C})$ and ${\cal
D}'_\geq({\cal C})$ in Section 
\ref{strange}. We consider the complexity of problems concerning small and 
large stable models in Sections \ref{small} and \ref{large}, respectively. 

\section{Preliminaries}
\label{prel-1}

We start by introducing some basic logic programming terminology. We
refer the reader to \cite{llo84,ap90} for a detailed treatment of the
subject. 

In the paper, we consider only the propositional case.
A logic program {\em clause} (or {\em rule}) is any expression $r$ 
of the form
\begin{eqnarray}
\label{cl-1}
r=\ \ p\lar q_1,\ldots,q_m,\n(s_1),\ldots,\n(s_n),
\end{eqnarray}
where $p$, $q_i$ and $s_i$ are propositional atoms. We call the atom $p$
the {\em head} of $r$ and we denote it by $h(r)$. Further, we call the 
set of atoms $\{q_1,\ldots,q_m,s_1,\ldots,s_n\}$ the {\em body} of $r$
and we denote it by $b(r)$. In addition, we distinguish the {\em 
positive body} of $r$, $\{q_1,\ldots,q_m\}$ ($b^+(r)$, in symbols), 
and the {\em  negative body} of $r$, $\{s_1,\ldots,s_n\}$ ($b^-(r)$,
in symbols).

A {\em logic program} is a collection of clauses. For a logic program 
$P$, by $\At(P)$ we denote the set of atoms that 
appear in $P$. If every clause in a logic program $P$ has an empty
negative body, we call $P$ a {\em Horn} program. If every clause in $P$
has an empty positive body, we call $P$ a {\em purely negative} program.

A clause $r$, given by (\ref{cl-1}), has a {\em propositional
interpretation} as an implication
\[
pr(r)=\ \ q_1\wedge\ldots\wedge q_m\wedge \neg s_1\wedge \ldots \wedge
\neg s_n
\Rightarrow p.
\]
Given a logic program $P$, by a {\em propositional interpretation} of 
$P$ we mean the propositional formula 
\[
pr(P) = \bigwedge\{pr(r)\colon r\in P\}.
\]
We say that a set of atoms $M$ is a {\em model} of a clause 
(\ref{cl-1}) if $M$ is a (propositional) model of the clause 
$pr(r)$. As usual, atoms in $M$ are interpreted as true, all 
other atoms are interpreted as false. A set of atoms $M\subseteq 
\At(P)$ is a {\em model} of a program $P$ if it is a model of 
the formula $pr(P)$. We emphasize the requirement $M\subseteq 
\At(P)$. In this paper, given a program $P$, we are interested 
only in the truth values of atoms that actually occur in $P$.

It is well known that every Horn program $P$ has a least model (with
respect to set inclusion). We will denote this model by $\lm(P)$.

Let $P$ be a logic program. Following \cite{cl78}, for every atom 
$p\in \At(P)$ we define a propositional formula $\comp(p)$ by
\[
\comp(p) = \ \ p\Leftrightarrow \bigvee\{c(r)\colon r\in P,\
h(r)=p\},
\]
where
\[
c(r) = \bigwedge\{q\colon q\in b^+(r)\}\wedge \bigwedge\{\neg
s\colon s\in b^-(r)\}.
\]
If for an atom $p\in \At(P)$ there are no rules with $p$ in the head, we
get an empty disjunction in the definition of $\comp(p)$, which we
interpret as a contradiction. 

We define the {\em program completion} (also referred to as the {\em 
Clark completion}) of $P$ as the propositional theory 
\[
\comp(P)=\{\comp(p)\colon p\in \At(P)\}.
\]
A set of atoms $M\subseteq \At(P)$ is called a {\em supported model} of 
$P$ if it is a model of the completion of $P$. It is easy to see that if 
$p$ does not appear as the head of a rule in $P$, $p$ is false in every 
supported model of $P$. It is also easy to see that each supported model 
of a program $P$ is a model of $P$ (the converse is not true in general). 

Given a logic program $P$ and a set of atoms $M$, we define the {\em
reduct} (also referred to as the {\em Gelfond-Lifschitz reduct}) of $P$ 
with respect to $M$ ($P^M$, in symbols) to be the logic program obtained 
from $P$ by
\begin{enumerate}
\item removing from $P$ each clause $r$ such that
$M\cap b^-(r)\not=\emptyset$ (we call such clauses {\em blocked by $M$}),
\item removing all negated atoms from the bodies of all the rules that
remain (that is, those rules that are not blocked by $M$).
\end{enumerate}
The reduct $P^M$ is a Horn program. Thus, it has a least model. We say
that $M$ is a {\em stable model} of $P$ if $M = \lm(P^M)$. Both the
notion of the reduct and that of a stable model were introduced in
\cite{gl88}. 

It follows directly from the definition that if $M$ 
is a stable model of a program $P$ then $M\subseteq
\At(P)$ and $M$ is a model of $P$. In fact, an even stronger 
property holds. It is well known that every stable model of 
a program $P$ is not only a model of $P$ --- it is a supported 
model of $P$. The converse does not hold in general. However, if a 
program $P$ is purely negative, then stable and supported models of 
$P$ coincide \cite{fag94}.

In our arguments we use fixed-parameter complexity results on 
problems to decide the existence of models of prescribed sizes for 
propositional formulas from some special classes. To describe these
problems we introduce additional terminology. First, given a
propositional theory $\Phi$, by $\At(\Phi)$ we denote the set of atoms
occurring in $\Phi$. As in the case of logic programming, we consider 
as models of a propositional theory $\Phi$ only those sets of atoms 
that are subsets of $\At(\Phi)$. Next, we define the following 
classes of formulas:
\begin{description}
\item[$\tn$:] the class of $t$-normalized formulas (if
$t=2$, these are simply CNF formulas)
\item[$\an_3$:] the class of all 2-normalized formulas whose every
clause is a disjunction of at most three literals (clearly, $\an_3$ is
a subclass of the class $\an$)
\item[$\tnm$:] the class of {\em monotone} $t$-normalized formulas,
that is, $t$-normalized formulas in which there are no occurrences of 
the negation operator
\item[$\tna$:] the class of {\em antimonotone} $t$-normalized formulas, 
that is, $t$-normalized formulas in which every atom is directly 
preceded by the negation operator.
\end{description}
Finally, we extend the notation ${\cal M}_\Delta({\cal C})$ and ${\cal
M}'_\Delta({\cal C})$, to the case when ${\cal C}$ stands for a class 
of propositional formulas. In this terminology, ${\cal M}'_=(\bnm)$ 
denotes the problem to decide whether a monotone 3-normalized 
formula 
$\Phi$ has a model in which exactly $k$ atoms are false. Similarly, 
${\cal M}_=(\tn)$ is simply another notation for the problem $\WS[t]$ 
that we discussed above. The following three theorems establish several 
complexity results that we will use later in the paper.

\begin{theorem}\label{tw0}
The problems ${\cal M}_=(\an)$, ${\cal M}_=(\anm)$, 
${\cal M}_{\leq}(\anm)$ and ${\cal M}'_=(\an)$ are all 
{\rm W[2]}-complete.
\end{theorem}
Proof: The first two statements, concerning the W[2]-completeness of 
${\cal M}_=(\an)$ and ${\cal M}_=(\anm)$, are proved in \cite{df97}. 

To prove the next statement, we will show that the problem ${\cal
M}_{\leq}(\anm)$ is equivalent to the problem ${\cal M}_{=}(\anm)$.
To this end, we first describe a reduction of ${\cal M}_{=}(\anm)$
to ${\cal M}_{\leq}(\anm)$. Let us consider a monotone 2-normalized 
formula $\Phi$ and an
integer $k$. If $k\leq |\At(\Phi)|$, we define $\Phi'=\Phi$ and $k'=k$. 
Otherwise, we define $\Phi=\{a\}$, where $a$ is a single-atom clause,
and $k'=0$. 

It is easy to see that $\Phi$ has a model with
exactly $k$ atoms if and only if $\Phi'$ has a model with at most $k'$
atoms. Indeed, let $M$ be a model of $\Phi$ with $k$ atoms. Since
$M\subseteq \At(\Phi)$, $k\leq |\At(\Phi)|$. Thus, $\Phi'=\Phi$ and 
$k'=k$. Consequently, $M$ is a model of $\Phi'$ and $|M|\leq k'$.

Conversely, let us consider a model $M$ of $\Phi'$ such that 
$|M|\leq k'$. Let us assume that $k > |\At(\Phi)|$. Then $\Phi'=
\{a\}$ and its only model is $\{a\}$. Since $k'=0$, this is a
contradiction with $|M|\leq k'$. Thus, $k\leq |\At(\Phi)|$ and we have
$\Phi'=\Phi$ and $k'=k$. It follows that there is a set $M'\subseteq
 \At(\Phi)$ such that $M\subseteq M'$
and $|M'|=k$. Since $\Phi$ is a monotone 2-normalized formula, a
superset of a model of $\Phi$ is also a model of $\Phi$. In particular, 
$M'$ is a model of $\Phi$ and it has exactly $k$ elements.

Given a pair $(\Phi,k)$, the pair $(\Phi',k')$ can clearly be 
constructed in time bounded by a polynomial in the size of $\Phi$. 
Thus, all the requirements of the Definition \ref{red} are satisfied.
Since $\Phi'$ is a monotone 2-normalized
formula, the problem ${\cal M}_{=}(\anm)$ is reducible to
the problem ${\cal M}_{\leq}(\anm)$.

The converse reduction can be constructed in a similar way. If $k\leq
|\At(\Phi)|$, we define $\Phi'=\Phi$ and $k'=k$. Otherwise, we define 
$\Phi'=\{a\}$, where $a$ is a single-atom clause,
and $k'=1$. It is easy to see that $\Phi$ has a model with
at most $k$ atoms if and only if $\Phi'$ has a model with exactly $k'$
atoms (a similar argument as before can be applied). Clearly, the pair
$(\Phi',k')$ can be constructed in time polynomial in the size of 
$\Phi$. Thus, the problem ${\cal M}_{\leq}(\anm)$ is reducible to
the problem ${\cal M}_{=}(\anm)$.
 
It follows that the problem 
${\cal M}_\leq(\anm)$ is equivalent to the problem ${\cal M}_{=}(\anm)$ 
which, as we already stated, is known to be W[2]-complete \cite{df97}.
Consequently, the problem ${\cal M}_{\leq}(\anm)$ is W[2]-complete. 

To prove the last statement of the theorem we reduce 
${\cal M}_=(\an)$ to ${\cal
M}'_=(\an)$ and conversely. Let us consider a 2-normalized formula
$\Phi = \bigwedge_{i=1}^m \bigvee_{j=1}^{m_i} x[i,j]$, where $x[i,j]$ 
are literals. We observe that $\Phi$ has a model of cardinality $k$ if 
and only if a related formula ${\bar \Phi}=\bigwedge_{i=1}^m
\bigvee_{j=1}^{m_i} {\bar x[i,j]}$, obtained from $\Phi$ by replacing 
every negative literal $\neg x$ by a new atom ${\bar x}$ and every positive 
literal $x$ by a negated atom $\neg{\bar x}$, has a model of cardinality 
$|\At({\bar \Phi})|-k$. This
construction defines a reduction of ${\cal M}_=(\an)$ to 
${\cal M}'_=(\an)$. It is easy to see that this reduction satisfies all
the requirements of the definition of fixed-parameter reducibility.

A reduction of ${\cal M}'_=(\an)$ to ${\cal M}_=(\an)$ can be 
constructed in a similar way. Since the problem 
${\cal M}_=(\an)$ is W[2]-complete, so is the problem 
${\cal M}'_=(\an)$. \hfill $\Box$

In the proof of Theorem \ref{tw0}, we presented several reductions
and observed that they satisfy all the requirements
specified in Definition \ref{red} of fixed-parameter reducibility. 
Throughout the paper we prove our complexity results by constructing
reductions from one problem to another. In most cases, we only verify 
the condition (\ref{red-4}) of the definition which, usually, is the
only non-trivial part of the proof. Checking that the remaining conditions 
hold is straightforward and we leave these details out.

\begin{theorem}\label{t-l}
The problems ${\cal M}_=(\an_3)$, ${\cal M}_=(\ana)$, 
${\cal M}'_=(\anm)$ and ${\cal M}'_\geq(\anm)$ are {\rm W[1]}-complete. 
\end{theorem}
Proof: The assertions concerning the first two problems are proved in 
\cite{df97}.

Using the reductions described in the proof of the last statement of
Theorem \ref{tw0}, it is easy to show that the problems ${\cal
M}'_=(\anm)$ and ${\cal M}_=(\ana)$ are equivalent. Thus, the problem
${\cal M}'_=(\anm)$ is W[1]-complete. 

Let $\Phi$ be a monotone 2-normalized theory. Clearly, $\Phi$ has a 
model of size at most $|\At(\Phi)|-k$ if and only if it has a model of size
exactly $|\At(\Phi)|-k$. 
Thus, the problem ${\cal M}'_\geq(\anm)$ is equivalent to the problem 
${\cal M}'_=(\anm)$. We have just proved that this last problem is
W[1]-complete. Thus, the problem ${\cal M}'_\geq(\anm)$ is also
W[1]-complete. \hfill$\Box$

\begin{theorem}\label{tw0.5}
The problems ${\cal M}'_=(\bnm)$ and ${\cal M}'_{\leq}(\bn)$ are 
$W[3]$-complete.
\end{theorem}
Proof: The problems ${\cal M}_=(\bna)$ and ${\cal M}_{\leq}(\bn)$ are
W[3]-complete \cite{df97}. Let us now observe that the 
problems ${\cal M}'_=(\bnm)$ and ${\cal M}_=(\bna)$ are equivalent.
Similarly, the problems ${\cal M}'_{\leq}(\bn)$ and ${\cal M}_{\leq}(\bn)$
are equivalent. Both equivalences can be argued in a similar way to that 
we used in the proof of the last statement of Theorem \ref{tw0}. 
Thus, the theorem follows.\hfill$\Box$

We will now present some general results that imply that in many cases, 
problems with $\Delta = \mbox{``$\leq$''}$, concerning stable and
supported models, are not harder than the corresponding problems with 
$\Delta = \mbox{``$=$''}$.

For every integer $k$, $1\leq k$, we denote by $Y_k$ the set of
propositional variables $y_{i,j}$, where $i=1,2\ldots,k+1$, and
$j=1,2,\ldots,i$. Next, for each $i$ and $j$, where $1\leq i\leq k+1$
and $1\leq j\leq i$, we define a logic program clause $q_{i,j}$ by:
\[
y_{i,j} \leftarrow \n (y_{1,1}),\ldots, \n (y_{i-1,1}),\n (y_{i+1,1}),\ldots, 
                   \n (y_{k+1,1})
\]
(let us note that for every $i$, $1\leq i\leq k+1$, rules $q_{i,j}$, 
$1\leq j\leq i$, have the same body). 
We then define a logic program $Q_k$ by setting  
\[
Q_k = \{q_{i,j}\colon 1\leq i\leq k+1\ \ \mbox{and}\ \  1\leq j\leq
i\}.
\]

\begin{lemma}
\label{lem1}
For every $i$, $1\leq i\leq k+1$, the set $\{y_{i,1},y_{i,2},\ldots,
y_{i,i}\}$ is a stable model (supported model) of $Q_k$. Moreover, $Q_k$
has no other stable models (supported models).
\end{lemma}
Proof: Let us consider any integer $i$ such that $1\leq i\leq k+1$.
We define $M = \{y_{i,1},y_{i,2},\ldots, y_{i,i}\}$. Since $y_{i,1}$
appears negated in the body of every rule $q_{i',j}$ of $Q_k$, with
$i'\not=i$ and $1\leq j\leq i'$, none of these rules contributes to 
the Gelfond-Lifschitz reduct of $Q_k$ with respect to $M$. On the other 
hand, no atom of $M$ appears negated in the bodies of the rules $q_{i,j}$, 
$1\leq j\leq i$. Thus, the Gelfond-Lifschitz reduct of $Q_k$ with
respect to $M$ consists of the rules
\[
y_{i,j}\leftarrow
\]
for $j=1,2,\ldots,i$. Clearly, the least model of the reduct is $M$ and,
consequently, $M$ is a stable model of $Q_k$. 

Let us consider now an arbitrary stable model $M$ of $Q_k$. Since $Q_k$ has
nonempty stable models and since stable models are incomparable with
respect to inclusion \cite{mt93}, $M\not=\emptyset$. Let $y_{i,j}\in 
M$, for some $i$ and $j$ such that 
$1\leq i\leq k+1$ and $1\leq j\leq i$. Since $q_{i,j}$ is the only
rule of $Q_k$ with the head $y_{i,j}$, it follows that its body is
satisfied by $M$. Since all rules $q_{i,j}$, $1\leq j\leq i$, have 
the same body and since $M$ is a model of $Q_k$, the heads of all these
rules belong to $M$. Thus, $\{y_{i,1},y_{i,2},\ldots, y_{i,i}\}
\subseteq M$. We proved earlier that $\{y_{i,1},y_{i,2},\ldots,
y_{i,i}\}$ is a stable model of $Q_k$. Since stable models are
incomparable with respect to inclusion,
$M = \{y_{i,1},y_{i,2},\ldots, y_{i,i}\}$. This completes the
proof of the assertion for the case of stable models.

The program $Q_k$ is purely negative. Thus, its stable and supported 
models coincide \cite{fag94}. Consequently, the assertion follows for 
the case of supported models, as well. \hfill$\Box$ 

\begin{theorem}
\label{leq}
Let $P$ be a logic program and let $k$ be a non-negative integer. Let
$Y_k =\{y_{i,j}\colon i=1,2\ldots,k+1,\ j=1,2,\ldots,i\}$ 
be a set of atoms disjoint with $\At(P)$ and let $Q_k$ be the 
program constructed above. Then:
\begin{enumerate}
\item $P$ has a supported model (stable model) of cardinality 
at most $k$ if and only if $P\cup Q_k$ has a supported model
(stable model) of cardinality equal to $k+1$.
\item $P$ has a supported model (stable model) of cardinality 
at least $|\At(P)|-k$ if and only if $P\cup Q_k$ has a supported model
(stable model) of cardinality equal to $|\At(P\cup Q_k)|-k(k+3)/2$.
\end{enumerate}
\end{theorem}
Proof: First, we observe that since $Y_k\cap \At(P) =\emptyset$,
supported models (stable models) of $P\cup Q_k$ are precisely the sets
$M'\cup M''$, where $M'$ is a supported model (stable model) of $P$
and $M''$ is a supported model (stable model) of $Q_k$.

The proofs for parts (1) and (2) of the assertion are very similar. We 
provide here only the proof for part (2). 

Let us assume that $M$ is a supported model of $P$ of cardinality at 
least $|\At(P)|-k$. Then, $|M|=|\At(P)|-k+a$, for some $a$, $0\leq
a\leq k$.
Clearly, $i=(k+1)-a$ satisfies $1\leq i\leq k+1$ and $\{y_{i,1},y_{i,2},
\ldots, y_{i,i}\}$ is a supported model of $Q_k$. It follows that 
$M'=M\cup \{y_{i,1},y_{i,2},\ldots, y_{i,i}\}$ is a supported model of 
$P\cup Q_k$ and its cardinality is $|\At(P)|-k+a +i$. It is now easy to
see that
\[
|\At(Q_k)| = (k+1)(k+2)/2.
\]
Thus, we have that 
\begin{eqnarray*}
|M'| &=&|\At(P)|-k+a +i = |\At(P)|+1\\
     &=& |\At(P\cup Q_k)| - (k+1)(k+2)/2 +1 = |\At(P\cup Q_k)| - k(k+3)/2.
\end{eqnarray*}  

Conversely, let us assume that $M'$ is a supported model of $P\cup Q_k$ of
cardinality exactly $|\At(P\cup Q_k)|-k(k+3)/2$. It follows that $M'=
M\cup \{y_{i,1},y_{i,2},\ldots, y_{i,i}\}$, where $M$ is a supported
model of $P$ and $1\leq i\leq k+1$. Clearly,
\begin{eqnarray*}
|M| &=& |M'|-i = |\At(P\cup Q_k)|-k(k+3)/2 -i\\
     &=& |\At(P)| + (k+1)(k+2)/2 - k(k+3)/2 -i \geq |\At(P)|-k.
\end{eqnarray*}
This completes the argument for part (2) of the assertion for the case 
of supported models. The same reasoning works also for the case of
stable models because all auxiliary facts used in this reasoning
hold for stable models, too. \hfill $\Box$ 

The program $Q_k$ can be constructed in time bounded by a polynomial in
the size of $P$ and $k$. Thus, Theorem \ref{leq} has the following
corollary on the reducibility of some problems ${\cal D}_\leq({\cal C})$
to the respective problems ${\cal D}_=({\cal C})$.

\begin{corollary}\label{eq-neq}
For every class of logic programs $\cal C$ such that $\cal C$ is closed
under unions and ${\cal N}\subseteq
{\cal C}$, problems $\SPM_\leq({\cal C})$, $\STM_\leq({\cal C})$,
$\SPM'_\leq({\cal C})$ and $\STM'_\leq({\cal C})$ can be reduced to (are
not harder than) problems $\SPM_=({\cal C})$, $\STM_=({\cal C})$,
$\SPM'_=({\cal C})$ and $\STM'_=({\cal C})$, respectively.
\hfill$\Box$
\end{corollary}

\section{The problems ${\cal D}_\geq({\cal C})$ and ${\cal
D}'_\geq({\cal C})$}
\label{strange}

These problems ask about the existence of models with at least
$k$ true atoms (in the case of small bound problems) or with at least 
$k$ false atoms (for the large-bound problems). From the point of view 
of the fixed-parameter complexity, these problems (with one exception)
are not very interesting. Several of them remain NP-complete even if $k$ 
is fixed (in other words, fixing $k$ does not render them tractable). 
Others are ``easy'' --- they can be solved in polynomial time even 
{\em without} fixing $k$. The one exception, the problem 
${\cal M}'_\geq({\cal N})$, turns out to be W[1]-complete.

\begin{theorem}
\label{t-p}
The following parameterized problems are in {\rm P}: 
${\cal M}_\geq({\cal H})$, ${\cal M}_\geq({\cal N})$, 
${\cal M}_\geq({\cal A})$, $\SPM_\geq({\cal H})$, $\STM_\geq({\cal
H})$, ${\cal M}'_\geq({\cal H})$, $\SPM'_\geq({\cal H})$ and $\STM'_\geq({\cal
H})$.
\end{theorem}
Proof: (1) The problems ${\cal M}_\geq({\cal H})$, 
${\cal M}_\geq({\cal N})$ and 
${\cal M}_\geq({\cal A})$ are all in P. Indeed, if $Q$ is a logic program, 
the set of all atoms of $Q$ is a model of $Q$. Thus, if $|At(Q)|\geq k$, 
the answer (in each case) is YES. Otherwise, the answer is NO. Clearly, 
the question whether $|At(Q)|\geq k$ can be decided in polynomial time
(in the size of $Q$ and $k$).\\
(2) $\SPM_\geq({\cal H})$ is in P. To see this, we observe that there 
is a polynomial-time algorithm to compute the greatest supported model of 
a Horn program \cite{ave82}. A Horn program 
$Q$ has a supported model of 
size at least $k$ if and only if the greatest supported model of $Q$ has 
size at least $k$. Thus, the assertion follows.\\
(3) The problem $\STM_\geq({\cal H})$ is in P. Indeed, the least model 
of a Horn program $Q$ is the only stable model of $Q$. The least model
of a Horn program $Q$ can be computed in linear time \cite{dg84}. So, 
the assertion follows.\\
(4) The problems ${\cal M}'_\geq({\cal H})$, $\SPM'_\geq({\cal H})$ and
$\STM'_\geq({\cal H})$ are all in P. Indeed, a Horn logic program has
the least model which is also the least supported and the only stable
model of $Q$. Thus, in the case of each of these three problems, the
answer is YES if and only if the least model of $Q$ has size at most 
$|\At(Q)|-k$. Since the least model of $Q$ can be computed in linear
time, the three assertions of (4) follow. \hfill$\Box$

In contrast to the problems covered by Theorem \ref{t-p}, which are
solvable in polynomial time even if $k$ is {\em not} a part of the 
input, problems in the next group remain hard even if $k$ is fixed.

\begin{theorem}
Let $k$ be a fixed non-negative integer. The following {\em
fixed-parameter} problems are {\rm NP}-complete: 
$\SPM_\geq({\cal N},k)$, $\SPM_\geq({\cal A},k)$, 
$\STM_\geq({\cal N},k)$, $\STM_\geq({\cal A},k)$,
$\SPM'_\geq({\cal N},k)$, $\SPM'_\geq({\cal A},k)$, 
$\STM'_\geq({\cal N},k)$ and $\STM'_\geq({\cal A},k)$.
\end{theorem}
Proof: (1) The problems $\SPM_\geq({\cal N},k)$, $\SPM_\geq({\cal A},k)$,
$\STM_\geq({\cal N},k)$ and $\STM_\geq({\cal A},k)$ are all NP-complete. 
Clearly, all these problems are in NP. To prove their NP-hardness, we 
recall that the problems to decide whether a logic program has a supported 
(stable) model are NP-complete, even under the restriction to purely 
negative programs \cite{mt88}. Let $P$ be a logic program. Let $y_i$, 
$i=1,2,\ldots, k$, be atoms not appearing in $P$. We define
\[
P' = P\cup \{y_i\leftarrow \ \colon\  i=1,2,\ldots, k\}.
\]
Since $\At(P)\cap \{y_1,y_2,\ldots,y_k\} = \emptyset$, $P$ has a stable
(supported) model if and only if $P'$ has a stable (supported) model of
size at least $k$. Moreover, if $P\in {\cal N}$, then $P'\in {\cal N}$, 
as well. Thus, NP-hardness of the problems
$\SPM_\geq({\cal N},k)$, $\SPM_\geq({\cal A},k)$,
$\STM_\geq({\cal N},k)$ and $\STM_\geq({\cal A},k)$ follows.\\
(2) The problems $\SPM'_\geq({\cal N},k)$, $\SPM'_\geq({\cal A},k)$, 
$\STM'_\geq({\cal N},k)$ and $\STM'_\geq({\cal A},k)$ are all NP-complete. 
Clearly, all these problems are in NP. To prove their NP-hardness, we use 
(as in (1)) the fact that the problems to decide whether a logic program 
has a supported (stable) model are NP-complete (even under the restriction 
to purely negative programs). Let $P$ be a logic program and let 
$y_i$, $z_i$, $i=1,2, \ldots,k$ be atoms not appearing in $P$. We define
\[
P' = P\cup \{y_i\leftarrow \n(z_i); z_i\leftarrow \n(y_i) \colon i=1,2,
\ldots, k\}.
\]
The logic program $\{y_i\leftarrow \n(z_i); z_i\leftarrow \n(y_i)
\colon i=1,2, \ldots, k\}$ has $2^k$ stable models. Each of these
models has exactly $k$ elements (for each $i=1,2,\ldots, k$, it contains
either $y_i$ or $z_i$ but not both). Since $\At(P)\cap \{y_i,z_i\colon 
i=1,\ldots,k\} = \emptyset$, $P$ has 
a stable (supported) model if and only if $P'$ has a stable (supported) 
model of size at most $|\At(P')|-k$. Moreover, if $P\in {\cal N}$ then
$P'\in {\cal N}$, as well. Thus, NP-hardness of the $\SPM'_\geq({\cal 
N},k)$, $\SPM'_\geq({\cal A},k)$, $\STM'_\geq({\cal N},k)$ and 
$\STM'_\geq({\cal A},k)$ follows.  \hfill$\Box$

We will next study the problem ${\cal M}'_\geq({\cal A},k)$. It turns
out that it is NP-complete for all $k\geq 1$ and is trivially solvable
in polynomial time if $k=0$.
 
\begin{theorem}
The problem ${\cal M}'_\geq({\cal A},0)$ is in P. For every $k\geq 1$,
the problem ${\cal M}'_\geq({\cal A},k)$ is NP-complete.
\end{theorem}
Proof. The first part of the assertion is evident. The answer to the
problem ${\cal M}'_\geq({\cal A},0)$ is always YES. Indeed, for every
logic program $P$, the set $M=\At(P)$ is a model of $P$ and it satisfies
the inequality $|\At(P)| \geq |M|$. 

Let us now assume that $k\geq 1$. We will first consider the problem 
${\cal P}(k)$ to decide whether a 2-normalized (that is, CNF) formula 
$\Phi$ has a model of size at most $|\At(\Phi)| - k$ ($k$ is fixed and not 
a part of the input). 
This problem is NP-complete. It is clearly in NP. To show its NP-hardness, 
we will reduce to it the general CNF satisfiability problem. Let $\Psi$ 
be a CNF theory and let $y_i$, $1\leq i\leq k$, be atoms
not occurring in $\Psi$. Then $\Psi$ has a model if and only if $\Psi'
=\Psi\cup \{\neg y_i\colon i=1,2,\ldots, k\}$ has a model of size at most
$|\At(\Psi')| - k$. Hence, NP-completeness of the problem ${\cal
P}(k)$, where $k\geq 1$, follows.

Problem ${\cal M}'_\geq({\cal A},k)$ is clearly in NP. To prove
NP-hardness of ${\cal M}'_\geq({\cal A},k)$ we will reduce the problem 
${\cal P}(k)$ to it. Let $\Phi$ be a CNF theory. Let us assume that 
$\At(\Phi) = \{x_1,x_2,\ldots,x_n\}$. For each
clause $C = a_1\vee\ldots\vee a_p\vee\neg b_1\vee\ldots\vee\neg b_r$ of
$\Phi$ we define program clauses $r_{C,i}$, $1\leq i\leq n$:
\[
r_{C,i}=\ \ x_i \leftarrow b_1,\ldots,b_r,\n(a_1),\ldots,\n(a_p).
\]
Let $P_\Phi =\{r_{C,i}\colon C\in \Phi,\ i=1,\ldots, n\}$. 
Clearly, $\At(P_\Phi)= \{x_1,x_2,\ldots,x_n\}$ (that is, the formula 
$\Phi$ and the program $P_\Phi$ have the same atoms). 

Let $M$ be a model of $\Phi$ and let $C$ be a clause of $\Phi$. Since
$M$ satisfies $C$, $M$ does not satisfy the body of the rules
$r_{C,i}$, $1\leq i\leq n$. In other words, $M$ satisfies all
the rules $r_{C,i}$, $1\leq i\leq n$. Thus, if $M$ is a model of $\Phi$
then $M$ is a model of $P_\Phi$. Since $\At(\Phi)=\At(P_\Phi)$, it
follows that if $\Phi$ has a model of size at most $|\At(\Phi)| - k$ then
the program $P_\Phi$ has a model of size at most $|\At(P_\Phi)| - k$. 

Conversely, let us consider a model $M$ of $P_\Phi$ such
that $|M|\leq n - k$. Since $k\geq 1$, we have $|M|< n$.
Let us assume that there is a clause $C$ of $\Phi$ that is not satisfied 
by $M$. Then, the bodies of all program clauses $r_{C,i}$, $1\leq 
i\leq n$, are satisfied. Hence, $\{x_1,\ldots, x_n\}\subseteq M$ and
$|M|\geq |\At(P_\Phi)|=n$, a contradiction. It follows that $M$ is a model
of $\Phi$. 

Thus, indeed, the problem ${\cal P}(k)$ can be reduced to the problem
${\cal M}'_\geq({\cal A},k)$ and NP-hardness of ${\cal M}'_\geq({\cal
A},k)$ follows.  \hfill$\Box$

The only problem with $\Delta=\mbox{``$\geq$''}$ whose complexity is 
affected by fixing $k$ is ${\cal M}'_\geq({\cal N})$. 
Namely, we have the following result. 

\begin{theorem}
\label{t3.4}
The problem ${\cal M}'_\geq({\cal N})$ is {\rm W[1]}-complete.
\end{theorem}
Proof: Let us consider a monotone 2-normalized formula $\Phi$. In each 
clause $C=x_1\vee\ldots\vee x_k$ of $\Phi$ we pick an arbitrary atom, 
say $x_1$. We then define a logic program clause $r_C = x_1 \leftarrow 
\n(x_2),\ldots, \n(x_k)$. Finally, we define a logic program $P_\Phi 
=\{r_C\colon C\in \Phi\}$. Clearly, $P_\Phi$ is a purely negative 
program, it is built over the same set of atoms as $\Phi$ and it has 
the same models as $\Phi$. Similarly, for every purely negative program 
$P$, the 2-normalized theory $pr(P)$ is monotone. Moreover, the set of 
atoms of $pr(P)$ is the same as that of $P$, and $pr(P)$ and $P$ have 
the same models.

It follows that the problem ${\cal M}'_\geq({\cal N})$ is equivalent
to the problem ${\cal M}'_\geq(\anm)$). Thus, the assertion follows by
Theorem \ref{t-l}. \hfill$\Box$

\section{The case of small models}
\label{small}

In this section we deal with the problems ${\cal M}_\Delta({\cal C})$,
$\SPM_\Delta({\cal C})$ and $\STM_\Delta({\cal C})$ for $\Delta =$ 
``$=$'' and $\Delta =$ ``$\leq$''. Speaking informally, we are interested 
in the existence of models that are {\em small}, that is, contain no
more than some specified number of atoms. The problem $\STM_\leq({\cal A})$ 
was first studied in \cite{tru99j}. In that work, it was proved that 
the problem $\STM_\leq({\cal A})$ is W[2]-hard and belongs to the class W[3].
In this section we establish the exact location of the problem 
$\STM_\leq({\cal A})$ in the W hierarchy and obtain similar results for 
problems concerning the existence of models and supported models. 

\begin{theorem}
The problems ${\cal M}_\leq({\cal N})$, ${\cal M}_=({\cal N})$, ${\cal
M}_\leq({\cal A})$ and ${\cal M}_=({\cal A})$ are all 
{\rm W[2]}-complete. 
\end{theorem}
Proof: Since ${\cal N}\subseteq {\cal A}$, it is enough to prove that 
the problems ${\cal M}_\Delta({\cal N})$, $\Delta =$ ``$\leq$'' and
``$=$'', are W[2]-hard, and that the problems ${\cal M}_\Delta({\cal
A})$, $\Delta =$ ``$\leq$'' and ``$=$'', are in W[2].

Reasoning as in the proof of Theorem \ref{t3.4}, we argue that
the problems ${\cal M}_\Delta(\anm)$ can be reduced to the problems
${\cal M}_\Delta({\cal N})$, for $\Delta =$ ``$\leq$'' and ``$=$''.
Indeed, $M$ is a model of a monotone 2-normalized formula $\Phi$ if and 
only if $M$ is a model of the logic program $P_\Phi$, as defined
in the proof of Theorem \ref{t3.4}. Since $\Phi$ is a
monotone 2-normalized formula, $P_\Phi$ is a purely negative logic 
program. This establishes the reducibility. By Theorem \ref{tw0}, it
follows that the problems ${\cal M}_\leq({\cal N})$ and ${\cal M}_=({\cal 
N})$ are W[2]-hard. 

Since $M$ is a model of a logic program $P$ if and only if $M$ is a 
model of $pr(P)$, it follows that the problems ${\cal M}_\leq({\cal
A})$ and ${\cal M}_=({\cal A})$ can be reduced to the problems ${\cal
M}_\leq(\an)$ and ${\cal M}_=(\an)$, respectively. Hence, by Theorem 
\ref{tw0}, the problems ${\cal M}_\leq({\cal A})$ and 
${\cal M}_=({\cal A})$ are in W[2]. \hfill$\Box$ 

\begin{theorem}
\label{t4.2}
The problems ${\cal M}_\leq({\cal H})$, $\SPM_\leq({\cal H})$,
$\STM_\leq({\cal H})$ and $\STM_=({\cal H})$ are in {\rm P}.
\end{theorem}
Proof: A Horn logic program has a model (supported model, stable model)
of size at most $k$ if and only if its least model (which is also the
least supported model and the only stable model) has size at most $k$.  
The least model of a Horn program can be computed in linear time. Thus,
the problems ${\cal M}_\leq({\cal H})$, $\SPM_\leq({\cal H})$ and
$\STM_\leq({\cal H})$ are in P. Since the least model of a Horn program 
is the {\em unique} stable model of the program, it follows that also
the problem $\STM_=({\cal H})$ is in P.
\hfill$\Box$

We emphasize that $k$ is a part of the input for problems dealt with in
Theorem \ref{t4.2}. Thus, all these problems are solvable in polynomial
time even without fixing $k$.

\begin{theorem} 
\label{t4.3}
The problem ${\cal M}_=({\cal H})$ is {\rm W[1]}-complete.
\end{theorem}
Proof: We will first prove the hardness part. To this end, we will
reduce the problem ${\cal M}_=(\ana)$ to the problem ${\cal M}_=({\cal
H})$. Let $\Phi$ be an 
antimonotone 2-normalized formula and let $k$ be a non-negative 
integer. Let $a_0, \ldots, a_{k}$ be $k+1$ different atoms not 
occurring in $\Phi$. For each clause $C=\neg x_1\vee\ldots \vee 
\neg x_p$ of $\Phi$ we define a 
logic program rule $r_{C}$ by 
\[
r_{C}=\ \ a_0 \leftarrow x_1,\ldots,x_p.
\] 
We then define $P_\Phi$ by
\[
P_\Phi = \{r_{C}\colon C\in \Phi\}\cup \{a_i\leftarrow a_j\colon
i,j=0,1,\ldots,k,\ i\not=j\}.
\]
Let us assume that $M$ is a model of size $k$ of the program $P_\Phi$.
If for some $i$, $0\leq i\leq k$, $a_i\in M$ then
$\{a_0,\ldots,a_k\}\subseteq M$ and, consequently, $|M|>k$, a 
contradiction. Thus, $M$ does not contain any of the atoms $a_i$. 
Since $M$ satisfies all rules $r_{C}$ and since it consists of atoms 
of $\Phi$ only, $M$ is a model of $\Phi$ (indeed, the body of each rule
$r_{C}$ must be false so, consequently, each clause $C$ must be true). 
Similarly, one can show that if $M$ is a model of $\Phi$ then it is a 
model of $P_\Phi$. Thus, W[1]-hardness follows by Theorem \ref{t-l}.

To prove that the problem ${\cal M}_=({\cal H})$ is in the class W[1], 
we will reduce it to the problem ${\cal M}_=(\an_3)$. To this end, for 
every Horn program $P$ we will describe a 2-normalized formula $\Phi_P$,
with each clause consisting of no more than three literals, and such 
that $P$ has a model of size $k$ if and only if $\Phi_P$ has a model of 
size $(k+1)2^k +k$. Moreover, we will show that $\Phi_P$ can be constructed 
in time bounded by a polynomial in the size of $P$ (with the degree not
depending on $k$).

First, let us observe that without loss of generality we may restrict
our attention to Horn programs whose rules do not contain multiple
occurrences of the same atom in the body. Such occurrences can be
eliminated in time linear in the size of the program. Next, let us note
that under this restriction, a Horn program $P$ has a model of size $k$ 
if and only if the program $P'$, obtained from $P$ by removing all clauses
with bodies consisting of more than $k$ atoms, has a model of size
$k$. The program $P'$ can be constructed in time linear in the size
of $P$ and $k$.

Thus, we will describe the construction of the formula $\Phi_P$ only for 
Horn programs $P$ in which the body of every rule consists of no more 
than $k$ atoms. Let $P$ be such a program. We define
\[
{\cal B} = \{B\colon\ B\subseteq b(r),\ \mbox{for some $r\in P$}\}.
\]
For every set $B\in {\cal B}$ we introduce a new variable $u[B]$.
Further, for every atom $x$ in $P$ we introduce $2^k$ new atoms $x[i]$, 
$i=1,\ldots, 2^k$. 

We will now define several families of formulas. First, for every 
$x\in \At(P)$ and $i=1,\ldots, 2^k$ we define
\[
D(x,i)=\ \ x\Leftrightarrow x[i] \ \ \ \ \mbox{(or $(\neg x \vee x[i])
\wedge (x\vee\neg x[i])$)},
\]
and, for each set $B\in {\cal B}$ and for each $x\in B$, we define
\[
E(B,x)=\ \ x\wedge u[B\setminus\{x\}] \Rightarrow u[B] \ \ \ \  
\mbox{(or $\neg x\vee \neg  u[B\setminus\{x\}] \vee u[B]$)}.
\]
Next, for each set $B\in {\cal B}$ and for each $x\in B$ we define
\[
F(B,x)=\ \ u[B]\Rightarrow x \ \ \ \ \mbox{(or $\neg u[B]\vee x$)}.
\]
Finally, for each rule $r$ in $P$ we introduce a formula
\[
G(r)=\ \ u[b(r)]\Rightarrow h(r) \ \ \ \ \mbox{(or $\neg u[b(r)]\vee h(r)$)}.
\]

We define $\Phi_P$ to be the conjunction of all these formulas 
(more precisely, of their 2-normalized representations given in the 
parentheses)
and of the formula $u[\emptyset]$. Clearly, $\Phi_P$ is a formula from 
the class $\an_3$.  Further, since the body of each rule in $P$ has at 
most $k$ elements, the set $\cal B$ has no more than $|P|2^k$ elements, 
each of them of size at most $k$ ($|P|$ denotes the cardinality of $P$,
that is, the number of rules in $P$). Thus, $\Phi_P$ can be constructed in 
time bounded by a polynomial in the size of $P$, whose degree does not 
depend on $k$. 

Let us consider a model $M$ of $P$ such that $|M|=k$. We define
\[
M' = M\cup \{x[i]\colon x\in M, i=1,\ldots,2^k\}\cup \{u[B]\colon
B\subseteq M\}.
\]
The set $M'$ satisfies all formulas $D(x,i)$, $x\in \At(P)$,
$i=1,\ldots, 2^k$. In addition,
the formula $u[\emptyset]$ is also satisfied by $M'$ ($\emptyset\subseteq
M$ and so, $u[\emptyset]\in M'$). 

Let us consider a formula $E(B,x)$, for some $B\in {\cal B}$ and $x\in
B$. Let us assume that $x\wedge u[B\setminus\{x\}]$ is true in
$M'$. Then, $x\in M'$ and, since $x\in \At(P)$, $x\in M$.
Moreover, since $u[B\setminus\{x\}]\in M'$, $B\setminus \{x\}\subseteq M$. 
It follows that $B\subseteq M$ and, consequently, that $u[B]\in M'$. Thus,
$M'$ satisfies all ``$E$-formulas'' in $\Phi_P$. 

Next, let us consider a formula $F(B,x)$, where $B\in {\cal B}$ and
$x\in B$, and let us assume that $M'$ satisfies $u[B]$. It follows that
$B\subseteq M$. Consequently, $x\in M$. Since $M\subseteq M'$, $M'$
satisfies $x$ and so, $M'$ satisfies $F(B,x)$. 

Lastly, let us look at a formula $G(r)$, where $r\in P$. Let us assume
that $u[b(r)]\in M'$. Then, $b(r)\subseteq M$. Since $r$ is a Horn
clause and since $M$ is a model of $P$, it follows that $h(r)\in M$. 
Consequently, $h(r)\in M'$. Thus, $M'$ is a model of $G(r)$.

We proved that $M'$ is a model of $\Phi_P$. Moreover, it is easy to see
that $|M'|= k+k2^k+2^k=(k+1)2^k + k$. 

Conversely, let us assume that $M'$
is a model of $\Phi_P$ and that $|M'|= (k+1)2^k + k$. We set
$M = M'\cap \At(P)$. First, we will show that $M$ is a model of $P$.
 
Let us consider an arbitrary clause $r\in P$, say
\[
r = h \leftarrow b_1,\ldots, b_p,
\]
where $h$ and $b_i$, $1\leq i\leq p$, are atoms. Let us assume that
$\{b_1,\ldots,b_p\}\subseteq M$. We need to show that $h\in M$.

Since  $\{b_1,\ldots,b_p\}=b(r)$, the set $\{b_1,\ldots,b_p\}$ and 
all its subsets belong to $\cal B$. Thus, $\Phi_P$ contains formulas 
\[
E(\{b_1,\ldots,b_{i-1}\},b_i) = b_i\wedge u[\{b_1,\ldots,b_{i-1}\}]
\Rightarrow u[\{b_1,\ldots,b_{i-1},b_i\}],
\]
where $i=1,\ldots, p$. All these formulas are satisfied by $M'$. We also 
have $u[\emptyset] \in \Phi_P$. Consequently, $u[\emptyset]$ is satisfied 
by $M'$, as well. Since all atoms $b_i$, $1\leq i\leq p$, are also
satisfied by $M'$ (since $M\subseteq M'$), it follows that $u[\{b_1,\ldots,
b_p\}]$ is satisfied by $M'$.

The formula $G(r) = u[\{b_1,\ldots,b_p\}]\Rightarrow h$ belongs to 
$\Phi_P$. Thus, it is satisfied by $M'$. It follows that $h\in M'$.
Since $h\in \At(P)$, $h\in M$. Thus, $M$ is a model of $r$ and,
consequently, of the program $P$.

To complete the proof we have to show that $|M|=k$. Since $M'$ is a
model of $\Phi_P$, for every $x\in M$, $M'$ contains all atoms $x[i]$,
$1\leq i\leq 2^k$. Hence, if $|M|>k$ then 
$|M'| \geq |M|+|M|\times 2^k \geq (k+1)(1+2^k) > (k+1)2^k +k$, a
contradiction. 

So, we will assume that $|M|<k$. Let us consider an
atom $u[B]$, where $B\in {\cal B}$, such that $u[B] \in M'$. 
For every $x\in B$, $\Phi_P$ contains the rule $F(B,x)$. The set 
$M'$ is a model of $F(B,x)$. Thus, $x\in M'$ and, since 
$x\in \At(P)$, we have that $x\in M$. It follows that $B\subseteq M$. 
It is now easy to
see that the number of atoms of the form $u[B]$ that are true in $M'$
is smaller than $2^k$. Thus, $|M'| < |M|+|M|\times 2^k +2^k \leq 
(k-1)(1+2^k) +2^k < (k+1)2^k +k$, again a contradiction. Consequently,
$|M|=k$. 

It follows that the problem ${\cal M}_=({\cal H})$ can be reduced to
the problem ${\cal M}_=(\an_3)$. Thus, by Theorem \ref{t-l}, the
problem ${\cal M}_=({\cal H})$ is in the class W[1]. This completes
our argument. \hfill$\Box$

\begin{theorem}\label{tw1}
The problems $\STM_{\leq}({\cal N})$ and $\SPM_{\leq}({\cal N})$ are
$W[2]$-hard.
\end{theorem}
Proof:  Since stable and supported models of purely negative programs
coincide \cite{fag94}, we will show $W[2]$-hardness for stable models 
only. To this end, we will find a reduction of ${\cal M}_{\leq}(\anm)$ 
(which is $W[2]$-hard, see Theorem \ref{tw0}) to $\STM_{\leq}({\cal N})$. 

Let $\Phi$ be a monotone 2-normalized formula and let $\{x_1,\ldots,x_n\}$
be the set of atoms that occur in $\Phi$. We define a program $P_\Phi 
\in {\cal N}$ as follows. For every atom $x_j$, $j=1,\ldots ,n$, 
occurring in $\Phi$ we introduce $k$ new atoms $x_j[1],x_j[2],\ldots ,
x_j[k]$. For each of these atoms we include in $P_\Phi$ the following 
rule:
\[ 
r_{j,\ell} =\ \  x_j[\ell ]\leftarrow {\rm\bf not}(x_1[\ell ]),\ldots
,{\rm\bf not}(x_{j-1}[\ell ]),
{\rm\bf not}(x_{j+1}[\ell ]),\ldots ,{\rm\bf not}(x_n[\ell ]),
\]
$j=1,\ldots ,n,\ \ell =1,\ldots ,k$. Next, for each clause $C=x_{i_1}\vee 
\ldots  \vee x_{i_s}$ in $\Phi$, we introduce a new atom $f_C$ and 
include in $P_\Phi$ the rule:
\begin{eqnarray*}
r_C =\ \  f_C\leftarrow&&{\rm\bf not}(x_{i_1}[1]),\ldots ,{\rm\bf
not}(x_{i_1}[k]),\\
&&{\rm\bf not}(x_{i_2}[1]),\ldots , {\rm\bf not}(x_{i_2}[k]),\\
&&\ldots ,\\
&&{\rm\bf  not}(x_{i_s}[1]),\ldots ,{\rm\bf not}(x_{i_s}[k]), {\rm\bf
not}(f_C).
\end{eqnarray*}

We will show that $\Phi$ has a model of cardinality at most $k$ if and
only if $P_\Phi$ has a stable model of size at most $k$.

Let $M=\{ x_{t_1},x_{t_2},\ldots ,x_{t_m}\}$, $m\leq k$, be a model of
$\Phi$. We claim that
\[
M'=\{ x_{t_1}[1],\ x_{t_2}[2],\ldots ,x_{t_m}[m],x_{t_m}[m+1],\ldots
,x_{t_m}[k]\}
\]
is a stable model of $P_\Phi$. Let $C$ be a clause from $\Phi$. Since $M$ 
is a model of $\Phi$, $C$ contains an atom, say $x_{t_j}$, from $M$. Then, 
however, $j\leq m$ and $x_{t_j}[j]\in M'$. The atom $x_{t_j}[j]$ occurs 
negated in the 
body of the rule $r_C$. Thus, the rule $r_C$ does not contribute to the 
reduct $P_\Phi^{M'}$. In the same time, the rules $r_{j,\ell}$
contribute the following rules to the reduct: 
\[ 
x_{t_j}[j]\leftarrow,
\]
for $j=1,\ldots ,m$, and
\[ 
x_{t_m}[j]\leftarrow ,
\]
for $j=m+1,\ldots ,k$. Thus, $\lm(P_\Phi^{M'})=M'$ and,
consequently, $M'$ is a stable model of $P_\Phi$ of size $k$.

Conversely, let us assume that $P_\Phi$ has a stable model $M'$ of size at 
most $k$. The atoms $f_C$ cannot be in $M'$ and, if $x_j[\ell ]\in M'$, 
then $x_i[\ell ]\not\in M'$, for $i\not= j$. Moreover, if for every
$j$, $1\leq j\leq n$, $x_j[\ell] \not\in M'$, then the rule $r_{1,\ell}$
implies that $x_1[\ell ]\in M'$, a contradiction. Hence, for every 
$\ell =1,\ldots, k$, exactly one of the atoms $x_1[\ell ],\ldots ,x_n[\ell ]$ 
is in $M'$. Thus, all stable models of $P_\Phi$ are of the form $M'=\{
x_{t_1}[1],x_{t_2}[2],\ldots ,x_{t_k}[k]\}$, where the indices
$t_1,t_2,\ldots ,t_k$ are not necessarily pairwise distinct. Let 
$M=\{ x_{t_1},\ldots ,x_{t_k}\}$.
Clearly, $|M|\leq k$. Suppose $M$ is not a model of some clause 
$C=x_{i_1}\vee \ldots \vee x_{i_s}$. Then, none of the atoms
$x_{i_1},\ldots ,x_{i_s}$ is in $M$. Consequently none of the atoms
$x_{i_j}[\ell ]$, $j=1,\ldots ,s$, $\ell =1,\ldots ,k$, is in $M'$.
It follows that the rule $f_C\leftarrow$ is in the reduct $P_\Phi^{M'}$ 
and, so, $f_C\in M'$, a contradiction. Thus, $M$ is indeed a model of 
$\Phi$ of cardinality at most $k$.  

This completes the argument that ${\cal M}_\leq(\anm)$ can be reduced 
to $\STM_\leq({\cal N})$
and the assertion of the theorem follows by Theorem \ref{tw0}. 
\hfill $\Box$

Later in the paper we will need a stronger version of Theorem
\ref{tw1}. To state it, we need more terminology. We define 
${\cal N}_1$ to be the class of purely negative programs such 
that each atom occurs exactly once in the head of a rule. It is clear
that the program $P_\Phi$ constructed in the proof of the Theorem
\ref{tw1} belongs to the class ${\cal N}_1$. Thus, we obtain the
following result.

\begin{theorem}\label{rem}
The problems $\STM_{\leq}({\cal N}_1)$ and $\SPM_{\leq}({\cal N}_1)$ 
are $W[2]$-hard. \hfill$\Box$ 
\end{theorem} 

\begin{theorem}\label{tw2}
The problem $\SPM_=({\cal A})$ is in $W[2]$.
\end{theorem}
Proof: We will show a reduction of $\SPM_=({\cal A})$ to ${\cal M}_=(\an)$, 
which is in $W[2]$ by Theorem \ref{tw0}. Let $P$ be a logic program 
with atoms $x_1,\ldots ,x_n$. We can identify supported models of $P$ 
with models of its completion $\comp(P)$. The completion is
of the form $\comp(P)=\Phi_1\wedge \ldots \wedge \Phi_n$, where
\[
\Phi_i=x_i\Leftrightarrow \bigvee_{j=1}^{m_i}\bigwedge_{\ell =1}^{m_{ij}}
x[i,j,\ell ],
\]
$i=1,\ldots ,n$, and $x[i,j,\ell ]$ are literals. It can be
constructed in linear time in the size of the program $P$.

We will use $\comp(P)$ to define a formula $\Phi_P$.
The atoms of $\Phi_P$ are $x_1, \ldots, x_n$ and $u[i,j]$, $i=1,
\ldots ,n$, $j=1,\ldots ,m_i$. For $i=1,\ldots ,n$, let 
\[
G_i= x_i \Rightarrow \bigvee_{j=1}^{m_i}u[i,j], \ \ \ \ \ ({\rm or}\ \neg 
x_i\vee\bigvee_{j=1}^{m_i}u[i,j]),
\]
\[
G_i'= \bigvee_{j=1}^{m_i}u[i,j] \Rightarrow x_i, \ \ \ \ \ ({\rm or}\
\bigwedge_{j=1}^{m_i}(x_i\vee\neg u[i,j])), 
\]
\[
H_i=\bigwedge_{j=1}^{m_i-1}\bigwedge_{j'=j+1}^{m_i}(\neg u[i,j]\vee\neg
u[i,j']),\quad \mbox{for every $i$ such that} \ m_i\geq 2,
\]
\[
I_i=\bigwedge_{j=1}^{m_i}(u[i,j]\Rightarrow\bigwedge_{\ell =1}^{m_{ij}}
x[i,j,\ell])\ \ \ \ \ ({\rm or}\ \bigwedge_{j=1}^{m_i}
\bigwedge_{\ell =1}^{m_{ij}}(\neg u[i,j]\vee x[i,j,\ell ])),
\]
\[
J_i= \bigvee_{j=1}^{m_i}\bigwedge_{\ell =1}^{m_{ij}}x[i,j,\ell]
\Rightarrow x_i,\ \ \ \ \ ({\rm or}\ \bigwedge_{j=1}^{m_i}(x_i\vee
\bigvee_{\ell =1}^{m_{ij}}\neg x[i,j,\ell])).
\]

The formula $\Phi_P$ is a conjunction of the formulas written above
(of the formulas given in the parentheses, to be precise). Clearly,
$\Phi_P$ is a 2-normalized formula. We will show that $\comp(P)$ has a 
model of size $k$ (or equivalently, that $P$ has a supported model of
size $k$) if and only if $\Phi_P$ has a model of size $2k$.

Let $M=\{ x_{p_1},\ldots ,x_{p_k}\}$ be a model of $\comp(P)$. Then, for 
each $i=p_1,\ldots ,p_k$, there is $j$, $1\leq j\leq m_i$, such
that $M$ is a model of $\bigwedge_{\ell =1}^{m_{ij}}x[i,j,\ell ]$
(this is because $M$ is a model of every formula $\Phi_i$). We denote one 
such $j$ (an arbitrary one) by $j_i$. We claim that 
\[
M'=M\cup \{ u[i,j_i]:\ i=p_1,\ldots ,p_k\}
\]
is a model of $\Phi_P$.
Clearly, $G_i$ is true in $M'$ for every $i$, $1\leq i \leq n$.
If $x_i\not\in M$ then $u[i,j]\not\in M'$ for all $j=1,\ldots,
m_i$. Thus, $G_i'$ is satisfied by $M'$. Since for each $i$, $1\leq
i\leq n$, there is at most one $j$ such that $u[i,j]\in M'$, it follows
that every formula $H_i$ is true in $M'$. By the definition of $j_i$, if 
$u[i,j]\in M'$ then $j=j_i$ and $M'$ is a model of 
$\bigwedge_{\ell =1}^{m_{ij}}x[i,j,\ell]$. Hence, $I_i$ is satisfied by 
$M'$. Finally, all formulas $J_i$, $1\leq i\leq n$, are clearly
true in $M'$. Thus, $M'$ is a model of $\Phi_P$ of size $2k$.

Conversely, let $M'$ be a model of $\Phi_P$ such that $|M'|=2k$.
Let us assume that $M'$ contains exactly $s$ atoms $u[i,j]$. The 
clauses $H_i$ ensure that for each $i$, $M'$ contains at most one atom 
$u[i,j]$. Therefore, the set $M'\cap \{u[i,j]\colon i=1,\ldots,n\
j=1,\ldots, m_i\}$ is of the form $\{u[p_1,j_{p_1}],\ldots,
u[p_s,j_{p_s}]\}$, where $p_1<\ldots<p_s$. 

Since the conjunction of $G_i$ and $G_i'$ is equivalent to
$x_i\Leftrightarrow \bigvee_{j=1}^{m_i}u[i,j]$, it follows that exactly 
$s$ atoms $x_i$ belong to $M'$. Thus, $|M'|=2s=2k$ and $s=k$. It is now
easy to see that $M'$ is of the form $\{ x_{p_1},\ldots ,x_{p_k},
u[p_1,j_{p_1}],\ldots , u[p_k,j_{p_k}]\}$. 

We will now prove that for every $i$, $1\leq i\leq n$, the implication
\begin{eqnarray*}
x_i\Rightarrow \bigvee_{j=1}^{m_i}\bigwedge_{\ell =1}^{m_{ij}}x[i,j,\ell ]
\end{eqnarray*}
is true in $M'$. To this end, let us assume that $x_i$ is true in $M'$ 
(in other words, that $x_i\in M'$). Then, there is $j$, $1\leq j\leq m_i$, 
such that 
$u[i,j]\in M'$ (in fact, $i=p_t$ and $j=j_{p_t}$, for some $t$, $1\leq 
t\leq k$). Since the formula $I_{i}$ is true in $M'$, the formula 
$\bigwedge_{\ell =1}^{m_{ij}}x[i,j,\ell ]$ is true in $M'$. Thus,
the formula $\bigvee_{j=1}^{m_i}\bigwedge_{\ell =1}^{m_{ij}}x[i,j,\ell
]$ is true in $M'$, too.

Since for every $i$, $1\leq i\leq n$, the formula $J_i$ is true in
$M'$, it follows that all formulas $\Phi_i$ are true in $M'$.
Since the only atoms of $M'$ that appear in the formulas $\Phi_i$ are
the atoms $x_{p_1}\ldots,x_{p_k}$, it follows that $M =
\{x_{p_1}\ldots,x_{p_k}\}$ is a model of $\comp(P)=\Phi_1\wedge
\ldots\wedge \Phi_n$.

Thus, the problem $\SPM_=({\cal A})$ can be reduced to the problem
${\cal M}_=(\an)$, which completes the proof. \hfill $\Box$

\begin{theorem}\label{tw3}
The problem $\STM_=({\cal A})$ is in $W[2]$.
\end{theorem}
Proof: In \cite{tru99j}, it is shown that the problem $\STM_=({\cal A})$ 
can be reduced to the problem of existence of a model of size $k$ of a 
certain formula $\Phi$. This formula $\Phi$ is a conjunction of
formulas of the form
\[ 
x_i\Leftrightarrow \bigvee_{j=1}^{m_i}\bigwedge_{\ell =1}^{m_{ij}}
x[i,j,\ell],
\]
for $i=1,\ldots ,n$, where $\{ x_1,\ldots ,x_n\}$ is the set of atoms
of $\Phi$ and $x[i,j,\ell ]$ denote some literals over this set of
atoms. This theory is the Clark 
completion of a certain logic program $P$. Thus, we get a reduction of 
$\STM_=({\cal A})$ to $\SPM_=({\cal A})$. By Theorem \ref{tw2}, 
it follows that $\STM_=({\cal A})$ is in $W[2]$.  \hfill $\Box$

Theorems \ref{tw1}, \ref{tw2} and \ref{tw3}, and Corollary \ref{eq-neq} 
yield the following result.

\begin{corollary}\label{wn4}
The problems $\STM_{\leq}({\cal N})$, $\SPM_{\leq}({\cal N})$,
$\STM_{\leq}({\cal A})$,
$\SPM_{\leq}({\cal A})$, $\STM_=({\cal N})$,  $\SPM_=({\cal N})$,
$\STM_=({\cal A})$ and $\SPM_=({\cal A})$ are $W[2]$-complete. \hfill
$\Box$
\end{corollary}

Finally, in our last result of this section, we establish bounds on the
fixed-parameter complexity of the problem $\SPM_=({\cal H})$.

\begin{theorem}\label{tw14}
The problem $\SPM_=({\cal H})$ is $W[1]$-hard and belongs to $W[2]$.
\end{theorem}
Proof: Since $\cal H$ is a subclass of $\cal A$, it follows immediately
from Theorem \ref{tw2} that  $\SPM_=({\cal H})$ is in $W[2]$.
The $W[1]$-hardness can be proved in exactly the same way as for the
problem ${\cal M}_=({\cal H})$ (Theorem \ref{t4.3}), except that for every
atom $x$ of $\Phi$, we have to include the rule $x\leftarrow 
x$ in $P_\Phi$. 
\hfill $\Box$

\section{The case of large models}
\label{large}

In this section we deal with the problems ${\cal M}'_\Delta({\cal C})$,
$\SPM'_\Delta({\cal C})$ and $\STM'_\Delta({\cal C})$ for $\Delta =$
``$=$'' and $\Delta =$ ``$\leq$''. Speaking informally, we are now 
interested in the existence of models that are {\em large}, that is, 
models in which the number of false atoms is bounded from above by some
integer. 
\begin{theorem}
The problems ${\cal M}'_\leq({\cal H})$, $\SPM'_\leq({\cal H})$, 
$\STM'_\leq({\cal H})$, ${\cal M}'_\leq({\cal N})$ and 
${\cal M}'_\leq({\cal A})$ are in {\rm P}.
\end{theorem}
Proof: The problems ${\cal M}'_\leq({\cal C})$, where ${\cal C}={\cal H},
{\cal N}$ or ${\cal A}$, have always the answer YES (the set of all atoms 
is a model of any logic program). Hence, all these three problems are
trivially in P.

Next, we observe that there is a polynomial-time algorithm to compute 
the greatest supported model of a Horn program \cite{ave82}. 
Consequently, the problem $\SPM'_\leq({\cal H})$ is in P 
(there is a supported model in which no more than $k$ atoms are false 
if and only if no more than $k$ atoms are false in the greatest supported 
model). 
Finally, a Horn program has a unique stable model (its least model) that 
can be computed in polynomial time.
Hence, the problem $\STM'_\leq({\cal H})$ is also in P.
\hfill$\Box$

\begin{theorem}
The problem ${\cal M}'_=({\cal N})$ is {\rm W[1]}-complete.
\end{theorem}
Proof: It is easy to see that this problem is equivalent to the
problem ${\cal M}'_=(\anm)$ (the same reductions as those used in
Theorem \ref{t3.4} work). This latter problem is W[1]-complete 
(Theorem \ref{t-l}). Hence, the assertion follows. \hfill$\Box$

\begin{theorem}
The problems ${\cal M}'_=({\cal H})$ and ${\cal M}'_=({\cal A})$ are 
{\rm W[2]}-complete.
\end{theorem}
Proof: Both problems are clearly in W[2] (models of a logic program $P$
are models of the corresponding 2-normalized formula $pr(P)$). Since
${\cal H}\subseteq {\cal A}$, to complete the proof it is enough to show 
that the problem ${\cal M}'_=({\cal H})$ is W[2]-hard. To this end, we 
will reduce the problem ${\cal M}_=(\anm)$ to ${\cal M}'_=({\cal H})$. 

Let $\Phi$ be a monotone 2-normalized formula and let $k\geq 0$. Let 
$\{x_1,\ldots,x_n\}$ 
be the set of atoms of $\Phi$. We define a Horn program $P_\Phi$ 
corresponding to $\Phi$ as follows. We choose an atom $a$ not occurring 
in $\Phi$ and include in $P_\Phi$ all rules of the form $x_i \leftarrow 
a$, $i=1,2,\ldots, n$. Next, for each clause $C = x_{i_1}\vee\ldots 
\vee x_{i_p}$ of $\Phi$ we include in $P_\Phi$ the rule
\[
r_{C} =\ \ \ a\leftarrow x_{i_1},\ldots , x_{i_p}.
\]
We will show that $\Phi$ has a model of size $k$ if and only if
$P_\Phi$ has a model of size $|\At(P_\Phi)|-(k+1)=(n+1)-(k+1)=n-k$. 

Let $M$ be a model of $\Phi$ of size $k$. We define $M'=\{x_1,\ldots,x_n\}
\setminus M$. The set $M'$ has $n-k$ elements. Let us consider any clause 
$r_C\in P_\Phi$ of the form given above. Since $M$ satisfies $C$, there 
is $j$, $1\leq j\leq p$, such that $x_{i_j} \notin M'$. Thus, $M'$ is a 
model of $r_{C}$. Since $a\notin M'$, $M'$ satisfies all clauses $x_i 
\leftarrow a$. Hence, $M'$ is a model of $P_\Phi$. 

Conversely, let $M'$ be a model of $P_\Phi$ of size exactly $n-k$. If 
$a\in M'$ then $x_i\in M'$, for every $i$, $1\leq i\leq n$. Thus,
$|M'|=n+1 > n-k$, a contradiction. Consequently, we obtain that $a\notin 
M'$. Let $M =
\{x_1,\ldots,x_n\}\setminus M'$. Since $a\notin M'$, $|M|=k$. Moreover,
$M$ satisfies all clauses in $\Phi$. Indeed, let us assume that there is
a clause $C$ such that no atom of $C$ is in $M$. Then, all
atoms of $C$ are in $M'$. Since $M'$ satisfies $r_{C}$, $a\in M'$,
a contradiction. Now, the assertion follows by Theorem \ref{tw0}.
\hfill$\Box$

\begin{theorem}\label{tw6}
The problem $\SPM_='({\cal H})$ is $W[3]$-hard. 
\end{theorem}
Proof: We will reduce the problem ${\cal M}'_=(\bnm)$ (which is
W[3]-complete by Theorem \ref{tw0.5}) to the problem
$\SPM'_=({\cal H})$. Let 
\[
\Phi =\bigwedge_{i=1}^m\bigvee_{j=1}^{m_i}\bigwedge_{\ell =1}^{m_{ij}}
x[i,j,\ell ] 
\]
be a monotone 3-normalized formula, where $x[i,j,\ell ]$ are atoms.
Let us assume that $|\At(\Phi)|=n$.

We define a Horn program $P_\Phi$ as follows. Let $u[1],\ldots ,u[m], v[1],
\ldots , v[k+1]$ be new atoms not occurring in ${\Phi}$. First, for every
$x\in \At(\Phi)$, we include in $P_\Phi$ the rule
\[ 
x\leftarrow x.
\] 
Next, for every $i=1,\ldots,m$, we include in $P_\Phi$ $m_i$ rules 
\[ 
u[i]\leftarrow x[i,j,1],\ldots,x[i,j,m_{ij}],
\]
where $j=1,\ldots,m_i$. Finally, we include in $P_\Phi$ $k+1$ rules
\[ 
v[q]\leftarrow u[1],\ldots,u[m],
\]
where $q=1,\ldots,k+1$.

We will show that ${\Phi}$ has a model of cardinality $n-k$ if and 
only if the Horn program $P_\Phi$ has a supported model of cardinality 
$|At(P_\Phi)|-k=n+m+k+1-k=n+m+1$.

Let $M$ be a model of $\Phi$, $|M|=n-k$. It is easy to see that 
$M'=M\cup \{ u[1],\ldots ,u[m],v[1],\ldots,$ $v[k+1]\}$ is a supported 
model of $P_\Phi$ of cardinality $n+m+1$.

Conversely, let $M'$ be a supported model of $P_\Phi$ of cardinality 
$n+m+1$. Clearly $M'$ is a model of the Clark completion $\comp(P_\Phi)$ 
of $P_\Phi$. If $u[i] \not\in M'$, for some $i=1,\ldots ,m$, then 
$v[q]\not\in M'$, for every $q=1,\ldots ,k+1$, because $v[q]
\Leftrightarrow \bigwedge_{i=1}^m u[i]$ belongs to $\comp(P_\Phi)$. Hence, 
$|M'|\leq n+m-1$, a contradiction. Therefore $u[i]\in M'$, for every $i=1,
\ldots ,m$. Consequently, for every $q=1,\ldots ,k+1$, we have
$v[q]\in M'$. Let $M=M'\cap \At(\Phi)$. Clearly, $|M|=
n+m+1-m-(k+1)=n-k$. Moreover, $M$ is a model of each formula 
$\bigvee_{j=1}^{m_i}\bigwedge_{\ell =1}^{m_{ij}}x[i,j,\ell ]$, $i=1,
\ldots ,m$. Indeed, $M'$ is a model of the formula $u[i]\Leftrightarrow 
\bigvee_{j=1}^{m_i}\bigwedge_{\ell =1}^{m_{ij}}x[i,j,\ell ]$ belonging 
to $\comp(P_\Phi)$ and $u[i]\in M'$, for $i=1,\ldots ,m$. Hence, $M$ 
is a model of $\Phi$ of cardinality $n-k$. \hfill $\Box$

\begin{theorem}\label{prop7}
The problem $\SPM_='({\cal A})$ is in $W[3]$. 
\end{theorem}
Proof: Let $P$ be a logic program with atoms $x_1,\ldots ,x_n$. Its 
supported models coincide with models of the Clark completion $\comp(P)$
of $P$. 
The formulas of the Clark completion are of the form 
\[
x_i\Leftrightarrow \bigvee_{j=1}^{m_i}\bigwedge_{\ell =1}^{m_{ij}}
x[i,j,\ell ],
\]
where $i=1,\ldots ,n$ and $x[i,j,\ell ]$ are literals. It is a routine
task to check that the completion $\comp(P)$ can be converted into a 
3-normalized formula in a number of steps being a polynomial with respect 
to the size of the program $P$. Hence, $\SPM_='({\cal A})$ is in $W[3]$. 
\hfill $\Box$

\begin{corollary}\label{wn8}
$\SPM_='({\cal H})$ and $\SPM_='({\cal A})$ are  {\rm W[3]}-complete.
\hfill$\Box$
\end{corollary}

\begin{theorem} \label{tw9}
The problems $\STM_{\leq}'({\cal A})$ and $\STM_{=}'({\cal A})$ are 
{\rm W[3]}-hard. 
\end{theorem}
Proof: By Corollary \ref{eq-neq}, it suffices to show that 
$\STM_{\leq}'({\cal A})$ is W[3]-hard. We will reduce the problem
${\cal M}'_{\leq}(3N)$ to the problem $\STM_{\leq}'({\cal A})$.
Let
\[
\Phi =\bigwedge_{i=1}^m\bigvee_{j=1}^{m_i}\bigwedge_{\ell =1}^{m_{ij}}
x[i,j,\ell]
\]
be a 3-normalized formula, where $x[i,j,\ell ]$ are literals. 
Let $u[1],\ldots,u[m]$, $v[1],\ldots ,v[2k+1]$ be new atoms not occurring in 
$\Phi$. For each atom $x\in \At(\Phi)$, we introduce new atoms $x[s]$, 
$s=1,\ldots,k$. 

Let $P_\Phi$ be a logic program with the following rules:
\[
A(x,y,s)=x[s]\leftarrow {\bf not}(y[s]),\quad x,y\in \At(\Phi),
\ x\not=y,\ s=1,\ldots ,k,
\]
\[ 
B(x)=x\leftarrow x[1],x[2],\ldots ,x[k],\quad x\in\At(\Phi),
\]
\[ 
C(i,j)=u[i]\leftarrow x'[i,j,1],x'[i,j,2],\ldots ,x'[i,j,m_{ij}],\quad 
i=1,\ldots ,m,\ j=1,\ldots ,m_i,
\]
where
\[
x'[i,j,\ell ]=\left\{ \begin{array}{ll}
                      x & {\rm if}\ x[i,j,\ell ]=x \\
                      {\bf not}(x) & {\rm if}\ x[i,j,\ell ]=\neg x,
                      \end{array}
                 \right.
\]
and
\[ 
D(q)=v[q]\leftarrow u[1],u[2],\ldots ,u[m],\quad q=1,\ldots ,2k+1.
\]

Clearly, 
$|\At(P_\Phi)|=nk+n+m+2k+1$, where $n=|\At(\Phi)|$. We will show that 
$\Phi$ has a model of 
cardinality at least $n-k$ if and only if $P_\Phi$ has a stable model 
of cardinality at least $|\At(P_\Phi)|-2k=n(k+1)+m+1$.

Let $M=\At(\Phi)\setminus\{ x_{1},\ldots ,x_{k}\}$ be a model of $\Phi$, 
where $x_{1},\ldots ,x_{k}$ are some atoms from $\At(\Phi)$ that are 
not necessarily distinct. We claim that $M'=\At(P_\Phi)\setminus
\{ x_{1},\ldots ,x_{k},x_{1}[1],\ldots ,x_{k}[k]\}$ 
is a stable model of $P_\Phi$.

Let us notice that a rule $A(x,y,s)$ is not blocked by $M'$
if and only if $y=x_s$. Hence, the program 
$P_\Phi^{M'}$ consists of the rules:
\[
x[1]\leftarrow\quad,\ {\rm for}\ x\not=x_1,
\]
\[ 
x[2]\leftarrow\quad,\ {\rm for}\ x\not=x_2
\]
\[
\ldots
\]
\[ 
x[k]\leftarrow\quad ,\ {\rm for}\ x\not=x_k
\]
\[ 
x\leftarrow x[1],x[2],\ldots ,x[k],\quad x\in\At(\Phi)
\]
\[ 
v[q]\leftarrow u[1],u[2],\ldots ,u[m],\quad q=1,\ldots ,2k+1,
\]
and of some of the rules with heads $u[i]$. Let us 
suppose that every rule of $P_\Phi$ with head $u[i]$ contains
a negated atom $x\in M$ or a non-negated atom $x\not\in M$. 
Then, for every $j=1,\ldots ,m_i$ there exists 
$\ell$, $1\leq \ell\leq m_{ij}$ 
such that either $x[i,j,\ell ]=\neg x$ and $x\in M$, or $x[i,j,\ell ]=x$ 
and $x\not\in M$. Thus, $M$ is not a model of the formula 
$\bigvee_{j=1}^{m_i} \bigwedge_{\ell =1}^{m_{ij}}x[i,j,\ell ]$ and,
consequently, $M$ is not a model of $\Phi$, a contradiction. Hence, for 
every $i=1,\ldots ,m$, there is a rule with head $u[i]$ containing 
neither a negated atom $x\in M$ nor a non-negated atom $x\not\in M$. 
These rules also contribute to the reduct $P_\Phi^{M'}$.

All atoms $x[s]\not= x_{1}[1],x_{2}[2],\ldots ,x_{k}[k]$ are 
facts in $P_\Phi^{M'}$. Thus, they belong to $\lm(P_\Phi^{M'})$. 
Conversely, if $x[s]\in \lm(P_\Phi^{M'})$ then $x[s]\not= x_{1}[1],
x_{2}[2], \ldots ,x_{k}[k]$. Moreover, it is evident by rules $B(x)$ 
that $x\in \lm(P_\Phi^{M'})$ if and only if $x\not= x_1,x_2, \ldots,
x_k$. Hence, by the observations in the previous paragraph, $u[i]\in 
\lm(P_\Phi^{M'})$, for each $i=1,\ldots ,m$. Finally, $v[q]\in 
\lm(P_\Phi^{M'})$, $q=1,\ldots 2k+1$, because the rules $D(q)$ belong 
to the reduct $P_\Phi^{M'}$. Hence, $M'=\lm(P_\Phi^{M'})$ so $M'$ is a 
stable model of $P_\Phi$ and its cardinality is at least $n(k+1)+m+1$.

Conversely, let $M'$ be a stable model of $P_\Phi$ of size at least 
$|\At(P_\Phi)|-2k$. Clearly all atoms $v[q]$, $q=1,\ldots ,2k+1$, must 
be members of $M'$ and, consequently, $u[i]\in M'$, for $i=1,\ldots ,m$. 
Hence, for each $i=1,\ldots ,m$, there is a rule in $P_\Phi$
\[ 
u[i]\leftarrow x'[i,j,1],x'[i,j,2],\ldots ,x'[i,j,m_{ij}]
\]
such that $x'[i,j,\ell ]\in M'$ if $x'[i,j,\ell ]=x$, and $x'[i,j,\ell]
\not\in M'$ if $x'[i,j,\ell ]=\neg x$. Thus, $M'$ is a model of the 
formula $\bigvee_{j=1}^{m_i}\bigwedge_{\ell =1}^{m_{ij}}x[i,j,\ell ]$, 
for each $i=1,\ldots ,m$. Therefore $M=M'\cap \At(\Phi)$ is a model of 
$\Phi$.

It is a routine task to check that rules $A(x,y,s)$ and $B(x)$ imply 
that all stable models of $P_\Phi$ are of the form 
\[ 
\At(P_\Phi)\setminus\{ x_{1},x_{2},\ldots ,x_{k},x_{1}[1],x_{2}[2],
\ldots , x_{k}[k]\}
\] 
(where $x_1,x_2,\ldots ,x_k$ are not necessarily distinct). Hence, $|M|=
|M'\cap \At(\Phi)|\geq n-k$. We have reduced the problem 
${\cal M}'_\leq(\bn)$ to the problem $\STM'_{\leq}({\cal A})$. Thus, 
the assertion follows by Theorem \ref{tw0.5}. \hfill $\Box$

\begin{corollary}\label{wn10}
The problem $\SPM_{\leq}'({\cal A})$ is {\rm W[3]}-hard. 
\end{corollary}
Proof: A {\em positive cycle} in a logic program $P$ is a sequence
of rules $r_0, r_1, \ldots, r_n$ in $P$ such that for every
$i=0,1,\ldots, n-1$, $h({r_i})\in b^+({r_{i+1}})$ and $h({r_n})\in
b^+({r_{0}})$. It is easy to see that the program $P$ constructed in 
the proof of Theorem \ref{tw9} does not contain positive cycles. 
Therefore, by the Fages lemma \cite{fag94}, stable and supported models 
of $P$ coincide. Thus, the proof of Theorem \ref{tw9} applies in the 
case of supported models too. \hfill $\Box$

By Theorem \ref{prop7}, Corollary \ref{wn10} and Corollary
\ref{eq-neq} we get the following result. 

\begin{corollary}\label{wn11}
The problem $\SPM_{\leq}'({\cal A})$ is {\rm W[3]}-complete.
\hfill$\Box$
\end{corollary}

\begin{theorem}\label{tw12}
The problem $\SPM_='({\cal N})$ is in {\rm W[2]}. 
\end{theorem} 
Proof: We will reduce the problem $\SPM_='({\cal N})$ to ${\cal M}'_=(\an)$  
(which belongs to W[2] by Theorem \ref{tw0}).

Let us consider a purely negative program $P$ with $\At(P)=
\{x_1,\ldots, x_n\}$. Its completion consists
of formulas
\[ 
\Phi_i =\ \ x_i\Leftrightarrow 
\bigvee_{j=1}^{m_i}\bigwedge_{\ell =1}^{m_{ij}}\neg x[i,j,\ell ] 
\ ,\quad i=1,\ldots ,n,
\]
where $x[i,j,\ell ]\in \At(P)$.

For each $x_i\in At(P)$, we introduce new atoms $x_i[1],x_i[2],\ldots ,
x_i[2^k]$. Next, for each set $U(i,j)=\{ x[i,j,\ell ]:\ \ell =1,
\ldots ,m_{ij}\}$ we define a new atom $u[i,j]$. Finally, we introduce  
yet another set of new atoms: $z[1],\ldots ,z[2^k]$. 

Let us consider the following formulas:
\[
A(i,t) =\ \  x_i\Leftrightarrow x_i[t],\quad i=1,\ldots ,n,\ 
t=1,\ldots ,2^k,
\]
\[ 
B(x,i,j) =\ \  x\Rightarrow u[i,j],\quad x\in U(i,j),\ 
i=1,\ldots ,n,\ j=1,\ldots ,m_i, 
\]
\[ 
C(i) =\ \  x_i\Rightarrow \bigvee_{j=1}^{m_i}\neg u[i,j],\quad i=1,\ldots ,n,
\]
\[ 
D(i) =\ \  x_i\Leftarrow \bigvee_{j=1}^{m_i}\bigwedge_{\ell =1}^{m_{ij}}\neg 
x[i,j,\ell ],\quad i=1,\ldots ,n,
\]
\[ 
E(t) =\ \ z[t]\Leftrightarrow z[t], \quad  t=1,\ldots 2^k.
\]

We define $\Phi_P$ to be the conjunction of the formulas listed above.
Since each of these formulas can be rewritten as a conjunction of
disjunctions, it is clear that without loss of generality we may assume
that $\Phi_P$ is a 2-normalized formula. Let us also note that the
number of atoms of $\Phi_P$ is given by the formula $|\At(\Phi_P)|=n(2^k+1)+
\sum_{i=1}^n m_i+2^k$.

We claim that $P$ has a supported model of size $n-k$ if and only if 
$\Phi_P$ has a model of size $|\At(\Phi_P)|-(k+1)2^k-k$. To prove 
it, we proceed as follows.

Let $\At(P)\setminus M$, where $M=\{ x_{i_1},\ldots ,x_{i_k}\}$, be a 
supported model of $P$ ($x_{i_1},\ldots, x_{i_k}$ are some $k$ distinct 
atoms of $P$). We denote by $q$ the number of subsets of 
$M$ different from all sets $U(i,j)$. We will show that $\At(\Phi_P)
\setminus M'$, where
\[
M'=M\cup \{ x_i[t]:\ x_i\in M,\ t=1,\ldots ,2^k\} \cup \{ u[i,j]:\ U(i,j)
\subseteq M\} \cup \{ z[1],\ldots ,z[q]\}
\]
is a model of $\Phi_P$. First, let us observe that $|M'|=k+k2^k+(2^k-q)
+q=(k+1)2^k+k$.

Clearly, by the definition of $M'$, $\At(\Phi_P)\setminus M'$ is a model 
of each formula $A(i,t)$. Let us consider a formula $B(x,i,j)$, for some 
$i,j$ such that $1\leq i\leq n$ and $1\leq j\leq m_i$, and for some $x
\in U(i,j)$. If $x\in \At(\Phi_P)\setminus M'$, then $x\not\in M$. 
It follows that $U(i,j) \not\subseteq M$. Consequently,
$u[i,j]\in \At(\Phi_P)\setminus M'$ and, so, $\At(\Phi_P)\setminus M'$ 
is a model of $B(x,i,j)$.

Next, let us consider a formula $C(i)$, for some $i$, $1\leq i\leq n$.
Further, let us assume that $x_i\in \At(\Phi_P)\setminus M'$. It follows 
that $x_i\in \At(P)\setminus M$. Since $\At(P)\setminus M$ is a supported 
model of $P$, $\At(P)\setminus M$ satisfies the formula $\Phi_i$. Thus,
there is $j$, $1\leq j\leq m_i$, such that for all $\ell =1, \ldots ,m_{ij}$, 
$x[i,j,\ell ]\in M$. Hence, $U(i,j)\subseteq M$ and, consequently,
$u[i,j]\notin\At(\Phi_P)\setminus M'$. Thus, $\At(\Phi_P)\setminus M'$ is a 
model of $C(i)$. 

Since $\At(P)\setminus M$ satisfies each formula $\Phi_i$, $1\leq i\leq 
n$, it is clear that $\At(\Phi_P)\setminus M'$ satisfies the formula 
$D(i)$. Since all formulas $E(t)$, $1\leq t\leq 2^k$, are tautologies,
$\At(\Phi_P)\setminus M'$ is a model of each of them, too. Thus,
$\At(\Phi_P)\setminus M'$ is a model of $\Phi_P$.
 
Conversely, let $\At(\Phi_P)\setminus M'$ be a model of $\Phi_P$, for 
some set $M'\subseteq \At(\Phi_P)$ such that $|M'|= (k+1)2^k+k$. Let 
$M=\At(P)\cap M'$. If $|M|>k$ then, since all formulas $A(i,t)$ hold in
$\At(\Phi_P)\setminus M'$, $|M'|\geq |M|(2^k+1)\geq (k+1)(2^k+1)>
(k+1)2^k+k$, a contradiction. Next, let us consider the case $|M|<k$ and 
let us assume that $u[i,j]\in M'$, for some $i$ and $j$ such that
$1\leq i\leq n$ and $1\leq j \leq m_i$. Since $\At(\Phi_P)\setminus M'$ is a
model of all formulas $B(x,i,j)$, where $x\in U(i,j)$, it follows that 
for every $x\in U(i,j)$, $x\in M$. Thus, $U(i,j)\subseteq M$ and, 
consequently, 
\[
|\{ u[i,j]:\ u[i,j]\in M'\} |=|\{ U(i,j):\ 
U(i,j)\subseteq M\} |\leq 2^{|M|}.
\]
Therefore, 
\[
|M'|\leq |M|(2^k+1)+2^{|M|}+2^k < (k+1)2^k+k-1<(k+1)2^k+k, 
\]
a contradiction again. Thus, $|M|=k$.

We will show that $\At(P)\setminus M$ is a supported model of $P$. To 
this end, we will prove that $\At(P)\setminus M$ is a model of all formulas
$\Phi_i$, $1\leq i\leq n$. Since $\At(\Phi_P)\setminus M'$ satisfies all 
formulas $D(i)$, $1\leq i\leq n$, and since all atoms appearing in these 
formulas belong to $\At(P)$, it follows that $\At(P)\setminus M$ satisfies 
all formulas $D(i)$, $1\leq i\leq n$. 

To show that $\At(P)\setminus M$ is a model of a formula $\Phi_i$, $1\leq 
i\leq n$, it is then sufficient to prove that $\At(P)\setminus M$ is a 
model of the implication 
\begin{eqnarray}\label{eq66}
x_i\Rightarrow \bigvee_{j=1}^{m_i} \bigwedge_{\ell =1}^{m_{ij}}
\neg x[i,j,\ell ]. 
\end{eqnarray}
Let $x_i\in \At(P)\setminus M$. Then, by the implication $C(i)$, which 
holds in $\At(\Phi_P)\setminus M'$, there exists $j$, $1\leq j\leq m_i$, 
such that $u[i,j]\in M'$. Using the implications $B(x,i,j)$ and reasoning 
as before, it is easy to show that $U(i,j)\subseteq M$. Thus, 
$\At(P)\setminus M$ is a model of $\bigwedge_{\ell =1}^{m_{ij}}\neg 
x[i,j,\ell ]$ and, consequently, of the implication (\ref{eq66}).
\hfill $\Box$

\medskip
A {\it kernel} of a digraph is an independent set $S$ of vertices 
(that is, a set of vertices with no edge with both the initial and 
terminal vertices in $S$) such that every vertex not in $S$ is a 
terminal vertex of some edge whose initial vertex is in $S$. 

Let us recall that ${\cal N}_1$ denotes the class of purely negative 
programs such that each atom occurs exactly once in the head of a rule. 
We define ${\cal N}_2$ to be the class of purely negative programs such 
that there is exactly one negated atom in the body of each rule.

Let $P\in {\cal N}_i$, $i=1,2$. We define $G(P)$ to be a digraph with the 
vertex set $At(P)$ and the edge set consisting of pairs $(y,x)$ such 
that there is a rule in $P$ with the head $x$ and ${\bf not}(y)$ in the 
body.

\begin{lemma}\label{lem5}
\begin{enumerate}
\item Let $P\in {\cal N}_1$. A set $S\subseteq At(P)$ is a stable 
model of $P$ if and only if $S$ is a kernel in $G(P)$.
\item Let $P\in {\cal N}_2$. A set $S\subseteq At(P)$ is a stable 
model of $P$ if and only if $At(P)\setminus S$ is a kernel in $G(P)$.
\end{enumerate}
\end{lemma}
Proof: (1) Let us assume that $S\subseteq \At(P)$ is a stable model of
a program $P\in {\cal N}_1$. For every $x\in\At(P)$, let us denote by 
$r_x$ the only rule of $P$ with $x$ as the head.

Let us consider a vertex $x\in S$. Then, $r_x$ is not blocked 
by $S$. Hence, for every $y$ in the body of $r_x$, $y\notin S$. In other 
words, for every $y$ such that $(y,x)$ is an edge of $G(P)$, $y\notin S$.
Thus, $S$ is an independent set. 

Next, let us consider a vertex $x\not\in S$. Then, $r_x$ is blocked by 
$S$. Consequently, there is $y$ in the body of $r_x$ such that $y\in S$.
In other words, there is an edge $(y,x)$ in $G(P)$ such that $y\in S$.

It follows that $S$ is a kernel of $G(P)$. The proof of the converse
implication is similar.

\noindent
(2) Let $S\subseteq At(P)$ be a stable model of a program $P\in {\cal 
N}_2$. Let us denote $S'=\At(P)\setminus S$. We will show that $S'$ is a
kernel of $G(P)$. Let $x\in S'$. Then $x\notin S$. Since $S$ is a stable 
model of $P$ and since $P\in {\cal N}_2$, it follows that every rule
$x\leftarrow\n(y)$ in $P$ is blocked by $S$ or, equivalently, that $y
\in S$. Consequently, for every edge $(y,x)$ in $G(P)$, if $x\in S'$,
then $y\notin S'$. Thus, $S'$ is an independent set.

Next, let us consider $x\notin S'$. Then, $x\in S$. Since $S$ is a stable 
model of $P$, there is a rule $x\leftarrow \n(y)$ in $P$ such that $y
\notin S$. It follows that $y\in S'$. Thus, for every $x\notin S'$, there 
is an edge $(y,x)$ in $G(P)$ such that $y\in S'$. Consequently, $S'$ is 
a kernel of $G(P)$. The proof of the converse statement is similar. 
\hfill $\Box$

\begin{theorem}\label{tw13}
The problems $\STM'_{\leq}({\cal N})$, $\STM'_=({\cal N})$, 
$\SPM'_{\leq}({\cal N})$ and $\SPM'_=({\cal N})$ are {\rm W[2]}-complete.
\end{theorem}
Proof: We will first reduce $\STM_{\leq}({\cal N}_1)$ to 
$\STM_{\leq}'({\cal N})$. Let $P\in {\cal N}_1$. We define $Q$ to be a 
program in ${\cal N}_2\subseteq {\cal N}$ such that $G(Q)=G(P)$. The 
program $Q$ is determined uniquely by the digraph $G(P)$. We will show 
that $P$ has a stable model of size at most $k$ if and only if $Q$ has 
a stable model of size at least $|At|-k$, where $\At$ is the set of atoms 
of both $Q$ and $P$. By Lemma \ref{lem5} $P$ has a stable model $S$ 
of size at most $k$ if and only if $S$ is a kernel of the digraph $G(P)$ 
of cardinality at most $k$. Lemma \ref{lem5} implies now that $G(Q)
=G(P)$ has a kernel $S$ of cardinality at most $k$ if and only if 
$\At\setminus S$ is a stable model of $Q$ of cardinality at least 
$|\At|-k$. 

It follows that the problem $\STM_{\leq}({\cal N}_1)$ can be reduced
to the problem $\STM_{\leq}'({\cal N})$. 
By Theorem \ref{rem} it follows that the problem 
$\STM'_{\leq}({\cal N})$ is W[2]-hard. Since stable and supported 
models of purely negative programs coincide, $\SPM_{\leq}'({\cal N})$ 
is W[2]-hard. Theorems \ref{tw12} and Corollary \ref{eq-neq} imply now
that both $\SPM_{\leq}'({\cal N})$ and $\SPM_='({\cal N})$ are 
W[2]-complete. The W[2]-completeness of the problems $\STM'_{\leq}({\cal 
N})$ and $\STM'_=({\cal N})$ follows again from the fact that stable and 
supported models coincide for purely negative programs. \hfill $\Box$

\section*{Acknowledgments}

This work was partially supported by the NSF grants CDA-9502645, 
IRI-9619233 and EPS-9874764.


\end{document}